\documentclass[prd,preprint,superscriptaddress,nofootinbib,longbibliography,aps]{revtex4-1}
\usepackage[utf8]{inputenc}
\usepackage{graphicx}
\usepackage{longtable}
\usepackage{amsmath}
\usepackage{color}
\usepackage{amssymb}
\usepackage{subfigure}
\usepackage{microtype}
\usepackage[linktoc=all]{hyperref}
\usepackage{cleveref}
\usepackage{stackengine}
\usepackage{color}
\usepackage{bm}
\usepackage{braket}
\usepackage{slashed}
\usepackage{tikz}
\usepackage{adjustbox}
\usepackage[compat=1.1.0]{tikz-feynman}

\definecolor{myblue}{rgb}{ 0.188, 0.478,0.858}

\newcommand{\mpl}{{M_{\rm {pl}}}}

\newcommand{\gsim}{\raisebox{-0.13cm}{~\shortstack{$>$ \\[-0.07cm]  $\sim$}}~}

\begin{document}
\preprint{Imperial/TP/2020/MC/01}

\title{An Effective Field Theory for Binary Cosmic Strings}

\author{Mariana Carrillo Gonzalez}
\email[Email: ]{m.carrillo-gonzalez@imperial.ac.uk}
\affiliation{Center for Particle Cosmology, Department of Physics and Astronomy, University of Pennsylvania 209 S. 33rd St., Philadelphia, PA 19104, USA}
\affiliation{Theoretical Physics, Blackett Laboratory, Imperial College, London, SW7 2AZ, U.K}
\author{Qiuyue Liang}
\email[Email: ]{qyliang@sas.upenn.edu}
\affiliation{Center for Particle Cosmology, Department of Physics and Astronomy, University of Pennsylvania 209 S. 33rd St., Philadelphia, PA 19104, USA}
\author{Mark Trodden}
\email[Email: ]{trodden@physics.upenn.edu}
\affiliation{Center for Particle Cosmology, Department of Physics and Astronomy, University of Pennsylvania 209 S. 33rd St., Philadelphia, PA 19104, USA}

\begin{abstract}
	\noindent
We extend the effective field theory (EFT) formalism for gravitational radiation from a binary system of compact objects to the case of extended objects. In particular, we study the EFT for a binary system consisting of two infinitely-long cosmic strings with small velocity and small spatial substructure, or ``wiggles". The complexity of the system requires the introduction of two perturbative expansion parameters, constructed from the velocity and size of the wiggles, in contrast with the point particle case, for which a single parameter is sufficient. This further requires us to assign new power counting rules in the system. We integrate out the modes corresponding to potential gravitons, yielding an effective action for the radiation gravitons. We show that this action describes a changing quadrupole, sourced by the bending modes of the string, which in turn generates gravitational waves. We study the ultraviolet divergences in this description, and use them to obtain the classical renormalization group flow of the string tension in such a setting.
\end{abstract}
\maketitle

\tableofcontents
\newpage

\section{Introduction}
The remarkable observations of gravitational waves by the LIGO and Virgo collaborations \cite{Abbott:2016blz,TheLIGOScientific:2017qsa} have opened new possibilities for exploring the universe. Understanding these observations requires the use of different theoretical and numerical tools to reproduce the gravitational waveforms. For example, the merging event of two compact objects consists of three phases: inspiral, merger, and ringdown; where each phase can be described through different techniques. The highly non-linear merging phase, in particular, requires the use of numerical relativity \cite{Pretorius:2005gq,Campanelli:2005dd,Baker:2005vv}. Fortunately, these time-costly techniques can be supplemented away from the merging phase by the use of theoretical techniques such as the effective one-body (EOB) formalism \cite{Buonanno:1998gg,Buonanno:2000ef} and the self-force formalism \cite{Quinn:1996am,Mino:1996nk}. During the inspiral phase, perturbative methods such as post-Newtonian(PN)~\cite{Einstein:1938yz,Jaranowski:1997ky,Damour:1999cr,Damour:2001bu,Damour:2014jta}, post-Minkowskian (PM)~\cite{Kerr:1959zlt,Bertotti:1960wuq,Westpfahl:1979gu,Bel:1981be,Ledvinka:2008tk,Westpfahl:1985tsl,Damour:2016gwp,Damour:2017zjx}, and non-relativistic general relativity (NRGR) effective field theory \cite{Goldberger:2004jt} can be used. The NRGR formalism consists of considering the effective action of point-particles with gravitational interactions, with higher dimensional operators encoding the finite size effects of the compact object being described. This method has led to great progress in computing PN corrections, not only in the spinless case, but also for spinning compact objects \cite{Porto:2005ac}. Recently, amplitudes techniques combined with EFT methods have been used to streamline the calculations in the PM approach~\cite{Bern:2019crd,Bern:2019nnu,Bern:2020buy,Chung:2020rrz,Damgaard:2019lfh,Aoude:2020onz,Vines:2018gqi} and for some PN computations \cite{Vaidya:2014kza}. Other EFT formulations for gravitating rotating objects which tackle dissipative effects with a different approach have been proposed in \cite{Endlich:2015mke,Endlich:2016jgc,Delacretaz:2014oxa}. For recent reviews on these different methods see \cite{Blanchet:2013haa,Schafer:2018kuf,Levi:2018nxp,Porto:2016pyg,Barack:2018yly}. 

In this paper, we explore the challenges of generalizing the NRGR EFT technique beyond point masses by considering the simple formal extension to one-dimensional sources. To be precise, we provide an EFT formalism to calculate the gravitational radiation from an artificial system of binary infinitely-long cosmic strings living in a 3+1 dimensional spacetime. While we do not expect to find such a simple realization of cosmic strings in nature, the goal is to illuminate and solve some of the issues that arise when one goes beyond the simplest systems in full gravity set up. Notice that other codimension 2 objects have been previously analyzed in different contexts, for example in \cite{deRham:2007mcp,Goldberger:2001tn}, and the observed behavior is similar to our case, as one might expect for localized objects of the same codimension.

A key feature that allows us to construct an EFT is the large separation of scales involved in the phenomenon being described. At low energies this allows us, by carefully choosing the degrees of freedom, to ignore short-range details of the theory, and to focus on the relevant physical effects at long length scales. For non-relativistic sources, the velocity always serves as a natural expansion parameter. In the case of point particles without gravity, it is easy to expand the worldline of the source to whatever order of velocity is required. However, for systems including gravity,  short-range gravitons may contribute to the background potential, making perturbation theory considerably more complicated. One can describe this system by integrating out the long-range gravitons and obtaining an effective field theory for the short distance degrees of freedom. A successful EFT assigns an order in the expansion parameter (the velocity in the point-particle case) to each term contributing to the effective action, and one may then compute observables up to a desired precision by truncating the EFT at the appropriate order.

The story for the cosmic string system is more involved than the point-particle case. To describe the string action, we must introduce an additional variable, $\sigma$, together with the proper time $\tau$, to describe string worldsheet. For a non-relativistic infinitely-long cosmic string, the velocity is not the only natural small parameter in the theory - we must also consider the spatial bending of the string. It is therefore natural to expect the effective field theory to contain two expansion parameters. This is made more difficult by the fact that the string equation of motion implies that the velocity and bending should be at the same order. We will show that, from virial theorem, the tiny difference between the velocity and the
bending should serve as the second expansion parameter. This novel power counting rule may ultimately be relevant for other studies involving multiple expansion parameters.

The gravitational potential between two binary infinitely-long cosmic strings has been rarely discussed. It is known that the solution of Einstein equations with one infinitely-long straight cosmic string at the lowest order is locally-flat with a deficit angle. For a wiggly string, the stress-energy tensor obtains a small correction leading to a gravitational field that a point-like test particle would feel. However, it is unclear what a test string feels in such a background. Here, we systematically calculate the gravitational potential between two cosmic strings and determine the order at which it becomes non-zero. We also obtain the gravitational radiation arising from a binary string system by calculating the imaginary part of its self-energy diagram. Compared to the point-particle case, it is possible for the cosmic string to generate a time-varying quadrupole itself, inducing gravitational waves with typical frequencies of the inverse of its wiggle scale. We will show that this contribution is larger than the one arising from gravitational interactions between the strings. However, we do not such gravitational waves to be detectable in the near future, since the dominant source of gravitational waves from cosmic strings originates in highly relativistic string loops, which our EFT fails to describe.

The paper is organized as follows. In Section \ref{sec,2}, we discuss how to build the EFT for the binary strings and radiation gravitons. We then derive the power counting rules of the EFT from the generalized virial theorem for binary strings. At the end of Section \ref{sec,2}, we explain the scaling of the Feynman graphs in the full theory which allow us to construct our EFT. In Section \ref{sec,3}, we discuss the EFT for Non-Relativistic strings. We integrate out the potential graviton to obtain the EFT that only contains radiation gravitons, and use this to calculate the gravitational radiation. In Section \ref{sec,4}, we consider a static, straight, relativistic string and compute the corrections to its stress-energy tensor from gravitational interactions. As expected, this calculation leads to divergences at 1-loop which can be regularized and renormalized. The renormalization of the string's tension leads to a classical RG flow which can be used to match this EFT coefficient to a UV description. In the appendix, we show further details of integrals used in the calculations in the main body of the paper.

\section{Building the EFT}
\label{sec,2}
\subsection{General strategy to construct an EFT}
The basic principle underlying the EFT idea is that an approximate description of physics at long length scales should not require detailed knowledge of the small-scale structure of the theory. Crucial to constructing an EFT for non-relativistic strings, therefore, is an understanding of the different relevant length scales in the model. These are: the finite size of the string, $r_s$;  the orbital distance between the binary strings, $r_o$; the size of the spatial substructure (wiggles) $r_w$; and the distance from the strings to the gravitational wave detector, $r_d$. It is the large separations between these scales
\begin{equation}
r_s \ll r_w \ll r_o \ll  r_d \ ,
\end{equation}
that allow us to construct a reliable EFT \footnote{We restrict ourselves to wiggles much larger than the size of the  string so that we can consistently include their description in our EFT.}. The EFT procedure sacrifices the granularity afforded by a complete ultraviolet (UV) description in favor of a calculationally simpler theory, in which high-energy physics is integrated out and its broad effects are instead encoded in higher-dimension operators. This allows us to retain in the EFT only those degrees of freedom that are relevant for the scales of interest in the problem at hand.

In this case, we do not expect that the microphysical details of the core of the string will be important to the question of gravitational radiation, and so we hope to avoid the complications of a field-theoretic description of the string itself, at scales of order $r_s$ or smaller. This is already a considerable simplification, since on scales larger than $r_s$, we have a good description of the system in terms of the string and Einstein-Hilbert (EH) actions. Nevertheless, what we are really interested in are the effects that a single string or binary strings should yield near the detector at which the gravitational radiation is measured. Thus, we seek an EFT in which all the degrees of freedom describing scales smaller than $r_o$ and $r_w$ are integrated out.

The highly nonlinear structure of General Relativity (GR) leads to an infinite number of vertices for graviton self-interactions, with a corresponding degree of calculational complexity. These calculations can be simplified if instead of using Newton's constant as an expansion parameter, we are able to use a new expansion parameter tailored to the specific situation at hand. For example, in Non-Relativistic (NR) systems such as compact spherical objects (treated as point-particles), the gravitational dynamics can be computed~\cite{Goldberger:2004jt} using a perturbative expansion in their velocity $v$. This NR approximation is extremely effective, but it should be noted that the formalism is only valid during the binary inspiral phase. One important feature that allows for this expansion is the fact that the virial theorem relates the gravitational coupling $1/\mpl$ to the velocity $v$. This provides a suppression, not only of higher-loop graphs, but also of higher $n-$point vertices. One can then truncate the infinite number of graviton vertices at a given order in the velocity.

In this paper, for the analogous case of a binary system of orbiting infinite strings, we will consider an extended non-relativistic limit in which the infinitely-long strings have both small velocity and small bending along their lengths, and will make a further assumption that the difference of the squares of these quantities is parametrically smaller than either of them individually. These assumptions will allow us to satisfy an appropriate version of the virial theorem and will again allow us to construct an EFT in which only a finite number of graphs contribute to physical process at certain order in these expansion parameters. Similar to the point-particle case, we expect that this approximation is only valid during the binary inspiral phase. Additionally, we will consider the case of a single NR string which can radiate gravitons on its own due to its own substructure, or wiggles. We will show that the binary strings can contribute to the generation of gravitational radiation at a lower order in our expansion parameter than that for the single string.

With the NR effective action for the string in hand, we will then use it to derive two different results. We first compute the gravitational potential between the pair of parallel strings and then calculate the radiation power emitted by a single string.

\subsection{Potential versus Radiation Gravitons}

In this section, we identify and combine the ingredients necessary to construct an effective field theory for non-relativistic binary strings. Our starting point is the action for Einstein gravity coupled to an arbitrary number $N$ of strings
\begin{equation}
S = 2 \mpl^2 \int d^4 x \sqrt{-g}~ R + S_{\rm string}\ , \label{EHaction}
\end{equation}
where we have made the convenient but somewhat non-standard definition $\mpl^2 \equiv (32\pi G)^{-1}$, with $G$ representing Newton's constant. Here, the string action is given at leading order by the Nambu-Goto action\footnote{We use the $(~-~+~+~+)$ signature throughout the paper.}
\begin{equation}
S_{\rm string} =-\sum_{i=1}^{N} T_i \int d^4 x  \int d^2\sigma_i \sqrt{- \det{\gamma}} \ \delta(x^0- \tau_i)\delta(x^1 - \sigma_i) \delta^2(\mathbf{x}-\mathbf{x_i})~, \label{stringaction}
\end{equation}
where $\gamma_{\alpha\beta}$ is the string worldsheet metric, $\mathbf{x_i}$ is the position of the $i$th string, $T_i$ is its tension, and $\{\sigma^0,\sigma^1\}=\{\tau,\sigma\}$ the worldsheet coordinates. Note that bold quantities represent 2d vectors in the directions perpendicular to the string. It is worth noting at this early stage that, thanks to the reparametrization invariance of the string action, we are free choose to align the worldsheet coordinates with the $x_0-x_1$ plane.

When constructing an EFT, we initially include all the terms that are consistent with the symmetries of our theory in the Lagrangian. The Einstein-Hilbert action itself contains the graviton self-interactions, while Eq.\eqref{stringaction} is the leading order contribution to the string-graviton interactions. In principle, we may include higher order operators which encode the effects of the finite size $r_s$ of the string (see \cite{Anderson:1997ip} and  references therein). For example, we might include the next leading order correction to the string action, given by
\begin{equation}
S\supset r_s^2 \int d^2\sigma \sqrt{- \det{\gamma}} \ \mathcal{K}_{a\mu\nu}^2  \  ,
\end{equation}
which corresponds to a non-minimal coupling of the worldsheet with the spacetime through the extrinsic curvature $\mathcal{K}_{a\mu\nu}$. Here we use Greek letters for the 4d spacetime coordinates, and Latin letters $a,b,\dots$ for the spatial directions orthogonal to the string, namely, the $x^2$ and $x^3$ components. However, we will not take such higher order terms into account in our treatment. Similarly, we will only consider non-rotating strings by ignoring possible spin degrees  of freedom.

An ambitious approach might be to seek an EFT for the strings themselves by integrating out all graviton degrees of freedom. However, this proves remarkably complicated and not suited for a non-relativistic expansion. Instead, we will take advantage of the different scales in our setup to first construct an EFT for the strings coupled only to those gravitons that can be observed near a gravitational wave detector - the so-called {\it radiation gravitons}.  In order to do this, we begin by defining the relativistic graviton $h_{\mu\nu}$ in the usual way, as a perturbation about the flat background geometry, writing the full metric as
\begin{equation}
g_{\mu\nu}\equiv \eta_{\mu\nu}+h_{\mu\nu} \ .
\end{equation}
We then split the relativistic graviton into a {\it potential} contribution, $H_{\mu\nu}$, and a {\it radiation} field, $\bar h_{\mu\nu}$. In this language, the potential gravitons are responsible for mediating the interactions between the strings. They are never on-shell, and thus they cannot appear as external lines in Feynman diagrams. The time scale relevant for these gravitons is set by the strings' velocity $\bm v = \partial \bm x/\partial \tau$ and the strings' orbital distance. Similarly, the length scale in the direction parallel to the strings is set by the strings' bending $\bm x' = \partial \bm x/\partial \sigma$, where prime denotes derivative with respect to $\sigma$, and the strings' orbital distance. Following this reasoning, we see that the potential gravitons' 4-momenta scale as:
\begin{equation}
k^\mu_\text{pot} \sim \left(\frac{v}{r_o},\frac{x' }{r_o},\frac{1}{r_o} ,\frac{1}{r_o} \right) \ ,
\end{equation}
where we have defined $v\equiv |\bm v|$ and $x'\equiv |\bm x'|$. Thus, potential gravitons are always space-like and correspond to short-range gravitational mediators.

Radiation gravitons, on the other hand, are responsible for the gravitational radiation, and thus can reach all the way to gravitational wave detectors. Their 4-momentum scales as
\begin{equation}
k^\mu_\text{rad} \sim \left(\frac{v}{r},\frac{x'}{r},\frac{\sqrt{ v^2- x'^2}}{r},\frac{\sqrt{ v^2-  x'^2}}{r} \right) \ , \label{krad}
\end{equation}
where, in the case of binary strings, $r$ is the orbital scale $r_o$, and for the case of a single string, $r$ is the scale of the wiggles $r_w$.  Since these gravitons are on-shell, we require that $( x')^2  \leq  v^2$. 
The on-shell condition also tells us that the length scale in the direction perpendicular to the string is set by $ r/\sqrt{ v^2 - x'^2} $, rather than by by $r$ as in the case of the off-shell potential gravitons. This implies that the radiation wavelength scales like $ r/\sqrt{ v^2 - x'^2} $. Comparing this with those gravitons arising from spherical extended sources \cite{Goldberger:2004jt}, we see that in our case the modes have longer wavelengths.

Decomposing the graviton into potential and radiation modes allows us to construct an effective action for the strings and the radiation graviton by integrating out the potential graviton modes according to
\begin{equation}
  \label{eq,seff2}
  \exp{\left[i S_{\rm eff}[x^i,\bar h]\right]} = \int \mathcal{D} H_{\mu\nu} \exp \left\{i S_{\rm EH}[\bar h + H] + i S_{\rm string}[x^i,\bar h + H]  + i S_{\rm GF}[H]\right\} \ ,
\end{equation}
where $S_{\rm GF}$ is a gauge fixing term. To keep gauge invariance manifest in the effective action, we choose a gauge fixing term invariant under diffeomorphisms of the background metric $\bar{g}_{\mu\nu} =\eta_{\mu\nu} + \bar{h}_{\mu\nu} $. We will work in the harmonic gauge defined by the gauge fixing term
\begin{equation}
	\label{eq,GF}
  S_{\rm GF} =-\mpl^2 \int d^4 x \sqrt{- \bar g} \Gamma_\mu \Gamma^\mu \   , \quad  \Gamma_\mu = D_\alpha H^\alpha_{~\mu} - \frac{1}{2} D_\mu H^\alpha_{~\alpha}  \ ,
\end{equation}
with $D_\alpha$ the covariant derivative with respect to the metric $\bar{g}_{\mu\nu} $. Taking into account the gauge fixing term, the propagator for potential gravitons is then given by
\begin{equation}
  \braket{H_{\mu\nu} (x_1)H_{\alpha\beta}(x_2) } = D_F(x_1 -x_2)P_{\mu\nu,\alpha\beta} \ ,
\end{equation}
where the propagator is given by
\begin{equation}
D_F(x_1 -x_2) = - \int \frac{d^4 k}{(2\pi)^4} \frac{i}{k^2- i \epsilon} e^{-i k\cdot  (x_1 -x_2)} \ ,
\end{equation}
and the Lorentz structure is encoded in
\begin{equation}
P_{\mu\nu,\alpha\beta} = \frac{1}{2}\left(\eta_{\mu\alpha}\eta_{\nu\beta}+\eta_{\mu\beta}\eta_{\nu\alpha}-\eta_{\mu\nu}\eta_{\alpha\beta}\right) \footnote{In D dimensions,  $P_{\mu\nu,\alpha\beta} = \frac{1}{2}\left(\eta_{\mu\alpha}\eta_{\nu\beta}+\eta_{\mu\beta}\eta_{\nu\alpha}- \frac{2}{D-2}\eta_{\mu\nu}\eta_{\alpha\beta}\right)$. }\ .
\end{equation}
For potential gravitons, $k^2 = -k_0^2 + k_1^2+ \bm k^2$ is dominated by $\bm k^2 \sim 1/r_o^2$. Therefore, we can approximate the propagator as
\begin{equation}
  \label{eq,propagator}
    D_F(x_1 -x_2) \simeq \left(-\int \frac{d^2 \bm k}{(2\pi)^2}  \frac{i}{ \bm k^2}  e^{i \bm k \cdot (\bm x_1 -\bm x_2)} \right)\delta(x^0)\delta(x^1)\sim  \mathcal{O}\left(\left(\frac{r_o}{\bm v}\right)^{-1} \left(\frac{r_o}{\bm x'}\right)^{-1} \log\left(\frac{r_o}{r_\text{IR}}\right)\right) \ .
\end{equation}
where $x^0 \sim r_o/ v$, $x^1 \sim r_o/x'$, and $r_\text{IR}$ is an IR cutoff. Further details are explained in Appendix~\ref{int}. Here, we have neglected the epsilon prescription in the Feynman propagator since the potential gravitons never go on-shell. This logarithmic behavior is precisely what we expect for a co-dimension-2 object, and is in agreement with the results of~\cite{Goldberger:2001tn}.

Rather than performing the path integral to obtain the effective action $S_{\rm eff}[x^i,\bar h]$, we will instead follow a much simpler procedure commonly known as matching \cite{Rothstein:2003mp,Cohen:2019wxr,Penco:2020kvy}. This consists of computing the Feynman diagrams in the full theory and constructing the effective action that reproduces them. In the following section, we will establish power counting rules for all the elements involved in the string EFT. Armed with these rules, we will be able to understand the scaling of different terms contributing to the effective action as a function of two distinct quantities which will serve as the expansion parameters. This expansion will then allow us to truncate the result, taking into account only the finite number of graphs that contribute at a given order.

\subsection{Power counting rules}
An important feature of the string EFT is the existence of a relationship between the gravitational coupling and the dynamical scales of our system, namely the velocity and the wiggles of the strings. This relationship leads to a truncation of the Feynman diagrams contributing at a given order in the expansion parameters. In the following, we will carefully derive this relation and simultaneously find the correct expansion parameters for our EFT. Once the appropriate expansion parameters have been found, we will  show how to obtain the scaling of graphs involving potential gravitons. Similarly, we will show in detail how to treat the radiation gravitons in our EFT.

\subsubsection{Virial theorem for binary strings}
\label{powcountrules}

As in the case of a system of binary point particles, the virial theorem plays a crucial role in enabling us to relate the gravitational coupling to the string velocity, although in the present application things are somewhat more involved. For a stable system bound by potential forces, the virial theorem can be generalized as
\begin{equation}
\left\langle \int d \sigma \ {\bf p}^a \dot{\bf x}_a\right\rangle=- \left\langle \int d \sigma \ {\bf x}_a \dot{\bf p}^a\right\rangle \ ,
\end{equation}
where in our case the brackets denote a time average over the whole inspiral phase. Using the string equation of motion, we have
\begin{equation}
\int d\sigma\ {\bf x}_a \dot{\bf p}^a =\int d\sigma \ {\bf x}_a \partial_\tau\frac{\partial L}{\partial \dot{\bf x}_a } = \int d\sigma\  {\bf x}_a \left(\frac{\partial L}{\partial {\bf x}_a} - \partial_\sigma\frac{\partial L}{\partial {\bf x}_a'}\right) =  \int d\sigma \left( {\bf{x}}_a \frac{\partial L}{\partial \bf{x}_a}+ \frac{\partial L}{\partial x_a'} x_a'\right) \ ,
\end{equation}
where we have integrated by parts in the last step. Thus, the virial theorem takes the form,
\begin{equation}
	\label{eq,virial}
	\left\langle \int d \sigma\left(\frac{\partial L}{\partial \dot{\bf x}_a} \dot{\bf x}_a + \frac{\partial L}{\partial {\bf x}_a'} {\bf x}_a'\right) \right\rangle=- \left\langle \int d \sigma \  {\bf{x}}_a \frac{\partial L}{\partial {\bf x}_a}\right\rangle \ .
\end{equation}
This result differs from the point particle case in two ways. First, for a system of point particles, the LHS is proportional to the kinetic energy of the system, while in our string system, an extra term describing wiggles on the string is present. Second, for a power-law potential, the RHS is simply proportional to $V$, corresponding to the potential energy; but this is not the case here.

We now apply this result to the case of a test  string moving in the spacetime generated by a static string with wiggles. After averaging over the wiggles, the stress energy tensor of a wiggly string can be expressed in terms of an effective tension $\tilde T$ and an effective energy density $\tilde \mu$ (both of which depend on the original tension $T$).  This effective stress tensor is valid when we are interested in physics on scales much larger that those of the spatial variations in the string structure. In this limit, the metric around a heavy wiggly static string is given by  $g_{\mu\nu} = \eta_{\mu\nu} + h_{\mu\nu}$ with
\begin{equation}
  h_{00}= h_{11} = 4G(\tilde \mu - \tilde T) \log\left(\frac{r}{r_\text{IR}}\right) \sim 4 GT ~\bm x'^2_\text{heavy} \log\left(\frac{r}{r_\text{IR}}\right)~, \label{h00}
\end{equation}
\begin{equation}
  h_{22}= h_{33} = 4 G(\tilde \mu + \tilde T) \log\left(\frac{r}{r_\text{IR}}\right) \sim 4 GT (2- \bm x'^2_\text{heavy}) \log\left(\frac{r}{r_\text{IR}}\right)  ~,  \label{h22}
\end{equation}
where $\bm x'_\text{heavy}$ denotes the wiggles of the heavy string~\cite{Vilenkin:1990mz}. In the special case of  a straight string, $\tilde \mu = \tilde T = T$.

Consider a test string living in this spacetime, the worldsheet volume element can be expanded as
\begin{align}
  \label{eq:expansion1string}
\sqrt{- \det{\gamma}}&=\left[(1 - \bm v^2 - h_{00}  -\bm v^2  h_{22}) ( 1+ \bm x'^2  +  h_{11} +\bm x'^2 h_{22}) +( \bm  v\cdot \bm x'  + \bm v\cdot \bm x' h_{22}   )^2 \right]^{1/2}~\nonumber\\
   &\simeq \left[1- (\bm v^2 - \bm x'^2) - (\bm v^2 - \bm x'^2) h_{22}  -(\bm v^2+\bm x'^2) h_{00} -(\bm v^2+\bm x'^2) h_{00} h_{22} -h_{00}^2 \right]^{1/2} \  ,
\end{align}
where in the second line we have used Eqs. \eqref{h00} and \eqref{h22} and have assumed $(\bm v\cdot\bm  x')^2 \sim \bm v^2 \bm x{'^2}$ which holds under time averaging.  In order to apply the virial theorem in this setting, it is convenient at this stage to expand the energy order by order in four expansion parameters defined in the following way
\begin{equation}
\epsilon_v \sim v^2 \ll 1 \ , \quad \epsilon_\Delta\sim\Delta \equiv v^2-x'^2 \ll 1  \ , \quad \epsilon_{GT}\sim \frac{T}{\mpl^2} \ll 1 \ ,\quad -\epsilon_h  \sim  -\frac{T}{\mpl^2} \log\left(\frac{r}{r_\text{IR}}\right) \ll 1 \ .
\end{equation}
In terms of these expansion parameters we can categorize terms in the action as $v^2 \sim \mathcal{O} (\epsilon_v) $, $x'^2 \sim \mathcal{O} (\epsilon_{v}-\epsilon_\Delta)$, $h_{00} \sim \mathcal{O} (\epsilon_{h}(\epsilon_{v}-\epsilon_\Delta))$, and $h_{22} \sim \mathcal{O} (\epsilon_{h}(2-\epsilon_{v}+ \epsilon_\Delta))$, where we have assumed that $\mathcal{O}( x'^2_\text{heavy}) =\mathcal{O}(x'^2) $. Note that then
\begin{equation}
	\frac{\partial h_{22}}{\partial \bm x^a}  \bm x^a = \frac{\partial h_{22}}{\partial r} \frac{\partial r}{\partial \bm x^a} \bm x^a =   \frac{\partial h_{22}}{\partial r} r  \sim \mathcal{O} (\epsilon_{GT}(2-\epsilon_{v}+\epsilon_\Delta))
\end{equation}
and
\begin{equation*}
\frac{\partial h_{00}}{\partial  \bm x^a}  \bm x^a \sim \mathcal{O} (\epsilon_{GT}(\epsilon_{v}-\epsilon_\Delta)) \ .
\end{equation*}
Using these expressions, and that for the Lagrangian of the test string, we find that the integrand on the left-hand-side of Eq.\eqref{eq,virial} is of order
\begin{eqnarray}
\mathcal{O}\left(\frac{\partial L}{\partial \dot{\bf x}_a} \dot{\bf x}_a + \frac{\partial L}{\partial {\bf x}_a'} {\bf x}_a'\right) &\sim& 2\epsilon_h \epsilon_v^2 + \epsilon_\Delta  \ ,
\end{eqnarray}
and similarly the integrand on the right-hand-side of Eq.\eqref{eq,virial} is of order
\begin{eqnarray}
\mathcal{O}\left(-{\bf{x}}_a \frac{\partial L}{\partial {\bf x}_a}\right) &\sim & -\epsilon_{GT}\epsilon_v^2
\end{eqnarray}
Requiring that  these  expressions agree leads us to choose the following relationships among our expansion parameters
 \begin{equation}
 \epsilon_\Delta \sim - \epsilon_v^2 \epsilon_{h} \  , \quad  \epsilon_{GT}\lesssim -\epsilon_h \ . \label{rel}
 \end{equation}
 Note that, since $-\epsilon_{h}\ll1$, Eq.\eqref{rel}  then implies that $ \epsilon_\Delta \ll \epsilon_v^2$. Similarly, the second relation in Eq.\eqref{rel}  implies that $\left|\log\left(\frac{r}{r_\text{IR}}\right)\right| \gsim \mathcal{O}(1)$. The above relations are particularly important, since they allow us to relate the gravitational coupling to the velocity $v$ and to the quantity $\Delta$ via
\begin{equation}
	\label{eq,grav}
 \left| \frac{T}{\mpl^2} \log\left(\frac{r}{r_\text{IR}}\right) \right| \sim \frac{\Delta }{v^4}  \ll 1\  .
\end{equation}
This is essential for the success of our EFT formalism. When we integrate out the potential gravitons to obtain an effective action for the radiation gravitons, Eq.\eqref{eq,grav} allows us to make a consistent truncation of the action.
Together, Eq.\eqref{rel} and Eq.\eqref{eq,grav} demonstrate that our original four expansion parameters were redundant, and that we only need two parameters, $ v$ and $\Delta$.

\subsubsection{Scaling of graphs involving potential gravitons}
Following the techniques of regular quantum field theory, a convenient way to integrate out the potential gravitons to obtain the effective action for the radiation gravitons is through the matching procedure --- using the power of Feynman diagrams to organize the perturbation expansion in terms of the above power counting rules. For example, the two-string interaction term arising from performing the path integral in Eq.\eqref{eq,seff2} can be found by  computing and summing a series of diagrams, representative examples of which are shown in Fig.\ref{fig,potentialEFT}.  It is necessary to understand how different Feynman diagrams scale with the expansion parameters in order to truncate the  infinite series at a desired order in $\Delta$ and $v$.  The key elements required to do so are the scaling of the internal lines, of the matter vertex, and of the n-pt graviton vertex. An internal potential graviton line scales as $D_F(x_i -x_j)$ given by Eq.\eqref{eq,propagator}.
Each matter vertex with $n_1$ gravitons introduces a factor of $\int d^2 \sigma T \left(1/\mpl\right)^{n_1}\sim \frac{r^2}{ v x'}  \frac{T}{\mpl^{n_1}}$, and each $n_2$-point graviton vertex coming from the EH action introduces a factor of $\left(1/\mpl\right)^{n_2-2} \sum_i^{n_2} k_i^2 \  \delta^4(\sum_i^{n_2}  k_i^2) $. This condition reduces the number of momentum-space integrations coming from the internal lines by one, and further removes one factor of $\frac{\bm v \bm x'  }{r^2}$. 
\begin{table}
\begin{tabular}{ |c |c |c |c |  }

 \hline
 \multicolumn{4}{|c|}{Power counting rules} \\
 \hline
  \raisebox{30\height}{ \begin{tikzpicture}
  \begin{feynman}
   \vertex (a1);
    \vertex[right=1cm of a1] (a2);
     \vertex[right=1cm of a2] (a3);
        \diagram* {
      (a1)-- [dashed](a2)-- [dashed](a3) ,
        };
  \end{feynman}
\end{tikzpicture}}
&  \raisebox{-0.4\height}{   \begin{tikzpicture}
  \begin{feynman}
   \vertex (a1);
    \vertex[right=1cm of a1] (a2);
     \vertex[right=1cm of a2] (a3);
    \vertex[below=2.5 em of a2] (b1){$\scriptscriptstyle \cdots$} ;
    \vertex[left=0.6 cm of b1] (b2){$\scriptscriptstyle 1$} ;
\vertex[right =0.3 cm of b2] (b4){$\scriptscriptstyle 2$} ;
    \vertex[right=0.6 cm of b1] (b3){$\scriptscriptstyle {n_1}$} ;

      \diagram* {
      (a1)-- [double](a2)-- [double](a3) ,
      (a2)-- [dashed](b2) ,
      (a2)-- [dashed](b3) ,
       (a2)-- [dashed](b4) ,
        };
  \end{feynman}
\end{tikzpicture}} &
 \raisebox{-0.3\height}{ \begin{tikzpicture}\centering
  \begin{feynman}
   \vertex (a1){$\scriptscriptstyle 2$};
     \vertex[right=0.3cm of a1] (a5){$\scriptscriptstyle 3$};
    \vertex[right=0.6cm of a1] (a2){$\scriptscriptstyle \cdots$};
     \vertex[below=0.6cm of a2] (a3);
     \vertex[right=1.2cm of a1] (a4){$\scriptscriptstyle n_2$};
     \vertex[below=0.6cm of a3] (b1){$\scriptscriptstyle 1$};
   
      \diagram* {
      (a1)-- [dashed](a3)-- [dashed](a4) ,
        (a3)-- [dashed](b1),
         (a3)-- [dashed](a5),
         };
  \end{feynman}
\end{tikzpicture}}  &\raisebox{0.6\height}{ \begin{tikzpicture}
  \begin{feynman}
   \vertex (a1);
    \vertex[right=2cm of a1] (a2);
     
        \diagram* {
      (a1)-- [gluon](a2) ,
        };
  \end{feynman}
\end{tikzpicture}} \\
 \hline
$\frac{\bm v \bm x'}{r^2} \log\left(\frac{r}{r_{\rm IR}}\right)$  & $\frac{r^2}{\bm v \bm x'} \frac{T}{\mpl^{n_1}}$  &$\left(\frac{1}{\mpl}\right)^{n_2-2} \sum_i^{n_2} k_i^2 \  \delta^4(\sum_i^{n_2}  k_i^2) $&  $\frac{\Delta}{r^2}$\\
 \hline
 \end{tabular}
 \caption{\label{tab:table-name} Power counting rules for vertices and propagators. The double line represents a cosmic string, the dashed line a potential graviton, and the curly line a radiation graviton. Note that the radiation graviton propagator does not have a logarithmic divergence as explained below. }

\end{table}
\vspace{1cm}

As an example, consider the second graph in Fig.\ref{fig,potentialEFT}. Using the rules described above, this diagram scales as 
\begin{equation}
 \frac{T^3}{\mpl^4}\int  \! d\sigma_{1,2,3}^2 \int_{\bm k_{1,2,3}}\!\!\!\!\!\!\! e^{i \sum_{i=1}^3 \bm k_i\cdot \bm x_i} \left(\!\delta^2(\sigma_1 -\sigma_2 )\delta^2(\sigma_1 -\sigma_3 ) \  \delta^2\left(\sum_{i=1}^3 \bm k_i \right) \prod_{i=1}^3 \frac{1}{\bm k_i^2}  \times \left[\sum_{i=1}^3 (\bm  k_{i\mu }\bm k_{i\nu})\right]\right) \  ,
\end{equation}
where $\sigma_{1,2,3} $ denote the worldsheet coordinates for each of the three string-graviton vertices, and the factor of $\left[\sum_i^3 (\bm  k_{i\mu }\bm k_{i\nu})\right]$ arises from the expansion of the EH term $\partial^2 h h h $. The square brackets indicate that all Lorentz indices are contracted appropriately. In this expression, we have canceled a factor of $\delta^2(\sigma)\sim \frac{r^2}{ v x'}$ coming from the internal lines with a factor of $\delta(k^0)\delta(k^1)\sim\frac{ v x'}{r^2}$ coming from the vertex. Simplifying the above result we find
\begin{equation}
\frac{T^3}{\mpl^4}\int   d\sigma^2\log\left(\frac{r}{r_\text{IR}}\right) \sim \mathcal{O} \left( \frac{L}{v^2}\left(\frac{\Delta}{v^4}\right)^2 \right)\ ,
\end{equation}
where we have performed the momentum integral and defined the angular momentum of the string as $L \equiv \int d \sigma T \bm r \cdot \bm v \sim T r^2 v/ x'$. In a similar fashion, we can use the power counting rules to compare the diagrams in  Fig.\ref{fig,power}. The tree-level graph on the left-hand side is of order $ \frac{L}{v^2} \frac{\Delta}{v^4}$, while the right-hand side diagram, which includes a graviton loop, scales as $\left(\frac{\Delta}{v^4}\right)^2$. This  shows that diagrams at loop order are $\frac{1}{L} \frac{\Delta}{v^2}$ suppressed\footnote{This requires $L\gsim v^3$ which is easily satisfied, since we expect $L\gg1$. } and thus we neglect them in our analysis.
	
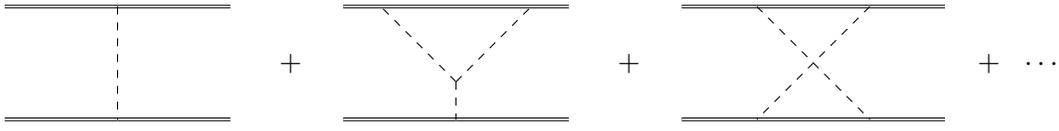
\begin{figure} 
\begin{tikzpicture}\centering
  \begin{feynman}
   \vertex (a1);
     \vertex[right=1.5cm of a1] (a2) ;
    \vertex[right=1.5cm of a2] (a3) ;
     \vertex[below=1.5cm of a1] (b1);
      \vertex[below=1.5cm of a2] (b2);
       \vertex[below=1.5cm of a3] (b3);

\vertex[right=0.8 cm of a3] (c1);
\vertex[below=0.5 cm of c1] (c2){$+ $} ;

\vertex[right=1.5cm of a3] (a4) ;
\vertex[right=1.5cm of a4] (a5) ;
\vertex[below=1 cm of a5] (a50) ;
     \vertex[left=1cm of a5] (a51) ;
     \vertex[right=1 cm of a5] (a52) ;
    \vertex[right=1.5cm of a5] (a6) ;
       \vertex[below=1.5cm of a4] (b4);
      \vertex[below=1.5cm of a5] (b5);
       \vertex[below=1.5cm of a6] (b6);
       
\vertex[right=0.8 cm of a6] (c3);
\vertex[below=0.5 cm of c3] (c4){$+ $} ;       

   \vertex[right=1.5cm of a6] (a7) ;
\vertex[right=1 cm of a7] (a8) ;
\vertex[right=1.5 cm of a8] (a9) ;
\vertex[right=1 cm of a9](a10) ;
\vertex[below=1.5 cm of a7] (b7) ;
\vertex[below=1.5 cm of a8] (b8) ;
\vertex[below=1.5 cm of a9] (b9) ;
\vertex[below=1.5 cm of a10] (b10) ;

\vertex[right=1.0 cm of a10] (c5);
\vertex[below=0.5 cm of c5] (c6){$+ ~~ \cdots $} ;     

      \diagram* {
      (a1)-- [double](a2) -- [double](a3)  ,
        (b1)-- [double](b2) -- [double](b3) ,
         (a2)-- [dashed](b2),
         };
\diagram* {
      (a4)-- [double](a6)  ,
        (b4)-- [double](b5) -- [double](b6) ,
         (a50)-- [dashed](b5),
         (a50)-- [dashed](a51),
	(a50)-- [dashed](a52),
         };
         \diagram* {
      (a7)-- [double](a10)  ,
        (b7)-- [double](b10) ,
         (a8)-- [dashed](b9),
         (a9)-- [dashed](b8),
         };
  \end{feynman}
  \end{tikzpicture}
  \caption{Representative Feynman diagrams contributing to the expansion of Eq.~\ref{eq,seff2} in the parameters $v$ and $\Delta$. Solid lines represent the source string, dashed lines represent the potential gravitons that are being integrated out.} \label{fig,potentialEFT}
\end{figure}

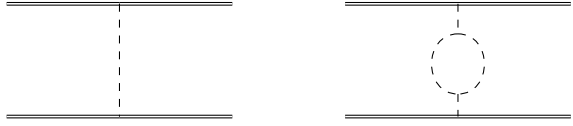
\begin{figure} 
\begin{tikzpicture}\centering
  \begin{feynman}
   \vertex (a1);
     \vertex[right=1.5cm of a1] (a2) ;
    \vertex[right=1.5cm of a2] (a3) ;
     \vertex[below=1.5cm of a1] (b1);
      \vertex[below=1.5cm of a2] (b2);
       \vertex[below=1.5cm of a3] (b3);

\vertex[right=1.5cm of a3] (a4) ;
\vertex[right=1.5cm of a4] (a5) ;
\vertex[below=0.4 cm of a5] (a50) ;
\vertex[below=1.2 cm of a5] (a51) ;
        \vertex[right=1 cm of a5] (a52) ;
    \vertex[right=1.5cm of a5] (a6) ;
       \vertex[below=1.5cm of a4] (b4);
      \vertex[below=1.5cm of a5] (b5);
       \vertex[below=1.5cm of a6] (b6);

      \diagram* {
      (a1)-- [double](a2) -- [double](a3)  ,
        (b1)-- [double](b2) -- [double](b3) ,
         (a2)-- [dashed](b2),
         };
\diagram* {
      (a4)-- [double](a6)  ,
        (b4)--[double](b6) ,
         (a5)-- [dashed](a50)--[dashed, half left](a51)--[dashed](b5),
         (a50)--[dashed, half right](a51)
         };        
           \end{feynman}
  \end{tikzpicture}
  \caption{Examples of diagrams contributing to the effective action with and without a graviton loop. The right diagram is $\frac{1}{L} \frac{\Delta}{v^2}$ suppressed relative to the left one.}  \label{fig,power}
\end{figure}

\subsubsection{Scaling of radiation gravitons}
Previously, we obtained the scaling of the potential graviton propagator by Fourier transforming to coordinate space. In that case, the 4-momentum was dominated by the directions perpendicular to the string, which led to a logarithmic behavior. As expected, in the case of the radiation graviton propagator the full 4-momentum is relevant. After Fourier transforming, we find that the coordinate-space radiation graviton scaling is 
\begin{equation}
\bar h_{\mu\nu}\sim \frac{\sqrt{ v^2 - x'^2}}{r}=\frac{\sqrt{\Delta}}{r} \ .
\end{equation}
As discussed in Eq.\eqref{krad}, $r$ is the orbital scale $r_o$ for binary strings and is the wiggle size $r_w$ for a single string. Thus, the scaling of radiation gravitons would be different in these two systems. If we assume $r_w \ll r_o$, then the scaling of the radiation graviton in the binary string system would be parametrically smaller than that in the single string system. 

Moreover, we can see that the elements of the EFT that involve radiation gravitons do not have a definite scaling in $v$ and $\Delta$ . For example, scattering amplitudes involving internal radiation gravitons behave like
\begin{equation}
  \int d^2\sigma_i d^2\sigma_j  \int \frac{d^4 k }{(2\pi)^4} \frac{-i}{k^2 -i \epsilon} e^{-ik^0(x_1^0- x_2^0) + i k^1 (x_1^1- x_2^1) + i \bm k^a (\bm x_{1a}- \bm x_{2a})}  \ .
\end{equation}
If we focus on the spatial part $e^{i \bm k^a \bm x_a} \sim 1+ i \bm k^a \bm x_a + \frac{1}{2} (i \bm k^a \bm x_a)^2 + \cdots$, we can see that the exponential contains an infinite series in powers of the  expansion parameters. A similar effect arises for the propagator of potential gravitons coupled to radiation gravitons, where there is  an infinite expansion in powers of ${\bf k}_\text{rad}/{\bf k}_\text{pot}\sim\sqrt{\Delta}$. To remedy these situations and to obtain an explicit scaling for the EFT action, we will perform a multipole expansion of the radiation graviton
\begin{equation}
\bar h_{\mu\nu}(x^0,x^1,\bm x^a)=\bar h_{\mu\nu}(x^0,x^1,\bm x_{CM}^a)+(\bm x^a-\bm x_{CM}^a) \partial_a \bar h_{\mu\nu}(x^0,x^1,\bm x_{CM}^a) + \cdots \ .
\end{equation}
In the following expressions, we will suppress the explicit functional dependence of the radiation graviton, but one should remember that it is evaluated at the center of mass $\bm x_{CM}$. Since the multipole expansion encodes the effects of the exponential expansion, we will now set $\bm x_{1a} = \bm x_{2a}$ at each order. With these power counting rules at hand, we are now ready to construct the EFT of non-relativistic strings as an example.

\section{The EFT for Non-Relativistic Strings}
\label{sec,3}
To construct the EFT for NR strings, we first need to integrate out the potential gravitons order by order. We begin by expanding the string Lagrangian for small velocity, wiggles, and metric perturbation. We then split the graviton into its potential and radiation parts, and discuss the gravitational potential between two cosmic strings and the gravitational radiation observed at a distant detector.

The worldsheet metric of the string is $\gamma_{\alpha\beta}  =g_{\mu\nu} \frac{\partial x^\mu}{\partial \sigma^\alpha}\frac{\partial x^\nu}{\partial \sigma^\beta}$, and we define $u^\mu \equiv \frac{\partial x^\mu}{\partial \tau } = (1,0,\bm v^a)~$ and  $w^\mu \equiv \frac{\partial x^\mu}{\partial \sigma } = (0,1,\bm x'^a) $, where $\bm v^a$ and $\bm x'^a$ stand for the string velocity and bending respectively.  From the discussion in Section~\ref{powcountrules} we know that the appropriate expansion parameters are $v$ and $\Delta$. However, when expanding the worldsheet volume element, we will encounter terms such as $\bm v^a h_{1a}$, which are not simple to classify according to these expansion parameters. For this reason, it is convenient to expand in factors of $v^2$ and $h$, and to perform the split $h = H + \bar h$ afterwards. This is useful for performing computations, but note that we will always return to the parameters $v$ and $\Delta$ in the effective action.

The worldsheet volume element can be expanded as,
\begin{align}
  \label{eq:expansion}
   \sqrt{- \det{\gamma}}& =  1+ \frac{1}{2}(-\bm v^2 +\bm x'^2 - h_{00}+  h_{11}   -  2 \bm v^a h_{0a}  + 2\bm x'^a h_{1a} -\bm v^a\bm v^b h_{ab} +\bm x'^a\bm x'^b h_{ab} \nonumber\\
&   -v^2 h_{11}  -\bm x'^2 h_{00} + 2\bm v^a \bm x'_a h_{01}
    -h_{00}h_{11}+ h_{01}^2 )-\frac{1}{8}\left[ 2(-\bm v^2 + \bm x'^2 )(- h_{00}+  h_{11}   )\right.\nonumber\\
    &\left.+ (- h_{00}+  h_{11}   )^2 \right] + \mathcal{O}(v^2 h^2)  \ ,
\end{align}
and then the action at lowest order is simply given by
\begin{equation}
  \label{eq,lowest}
S^\text{eff}_{L \frac{\Delta}{v^2} } =  \frac{1}{2} \sum_i  T_i  \int d^2\sigma_i  \left(\bm v_i^2-\bm x_i'^2\right) \ ,
\end{equation}
where $L \frac{\Delta}{v^2}$ denotes the order of the action  and $L$ is the angular momentum defined earlier. The leading order Equation of Motion (EoM) for $\bm x$ is then
\begin{equation}
  \label{eq,EoM}
  \ddot{\bm x} - \bm x'' =0 \ .
\end{equation}
A simple question is then; at which order do gravitational forces start to contribute for a slightly wiggly string with a small velocity? In the following section, we will show that the gravitational potential for binary strings first enters at order $L \Delta^2/v^6 $, which is $ {\Delta }/{v^4}$ smaller than the leading order contribution.

\subsection{The gravitational potential between binary strings}
We now focus on the contributions of potential gravitons --- $H_{\mu\nu} $--- in the expansion of  Eq.\eqref{eq:expansion}, and compute the gravitational potential between binary strings. The interaction between a string and a potential graviton at $\mathcal{O}(v^0 H)$ reads,
\begin{equation}
  S =  \frac{T}{2\mpl} \int d^2\sigma (H_{ 00}- H_{ 11}) \ .
\end{equation}
The vertex for this interaction is shown in Fig.\ref{fig,vertex1} (a) and the corresponding momentum-space Feynman rule is,
 \begin{equation}
   \label{eq,vertex1}
    \Gamma^{(0)}_{\mu\nu}= \frac{i ~T}{\mpl}  \int d^2\sigma ~ V^{(0)}_{\mu\nu}~,\quad
     V^{(0)}_{\mu\nu} =\frac{1}{2}\left(I_{00,\mu\nu}-I_{11,\mu\nu}\right) \ ,
 \end{equation}
where the tensor $I_{\mu\nu,\alpha\beta} \equiv \frac{1}{2} (\eta_{\mu\alpha}\eta_{\nu\beta} + \eta_{\mu\beta}\eta_{\nu\alpha})$ is symmetric in all of its indices and the superscript $(n)$ denotes the order in $v$. Using this, we can see that the diagram in Fig.\ref{fig,vertex1} (b) is,
\begin{equation}
  \text{Fig.\ref{fig,vertex1}~(b)}= \frac{ - ~T_1 T_2}{ \mpl^2}  \int d^2\sigma_{1} d^2\sigma_{2}  V^{(0)}_{\mu\nu}    V^{(0)}_{\alpha\beta}  \braket{H^{\mu\nu}H^{\alpha\beta}}\propto V^{(0)}_{\mu\nu} P^{\mu\nu,\alpha\beta} V^{(0)}_{\alpha\beta} = 0  \ ,\label{potzero}
\end{equation}
which vanishes automatically due to the symmetry of the tensor structure.

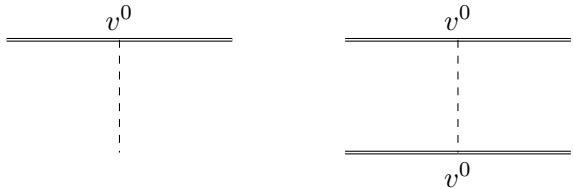
\begin{figure} 
\begin{tikzpicture}\centering
  \begin{feynman}
   \vertex (a1);
     \vertex[right=1.5cm of a1] (a2);
    \vertex[right=1.5cm of a2] (a3) ;
     \vertex[below=1.5cm of a2] (b2);

\vertex[right=1.5cm of a3] (a4) ;
\vertex[right=1.5cm of a4] (a5) ;
    \vertex[right=1.5cm of a5] (a6) ;
       \vertex[below=1.5cm of a4] (b4);
      \vertex[below=1.5cm of a5] (b5);
       \vertex[below=1.5cm of a6] (b6);

      \diagram* {
      (a1)-- [double, edge label=$ v^0$ ](a3)  ,
              (a2)-- [dashed](b2),
         };
\diagram* {
      (a4)-- [double, edge label=$ v^0$](a6)  ,
        (b4)--[double, edge label'=$ v^0$](b6) ,
         (a5)-- [dashed](b5),
                  };        
           \end{feynman}
  \end{tikzpicture}
  \caption{Left panel: the vertex associated with the Feynman rule in Eq.\eqref{eq,vertex1}. The solid line represents a NR string source and the dashed line is a potential graviton. Right panel: the leading order contribution to the potential energy between two strings arising from the vertex in Eq.\eqref{eq,vertex1}; this contribution vanishes, as shown in Eq.\eqref{potzero}.} \label{fig,vertex1}
\end{figure}

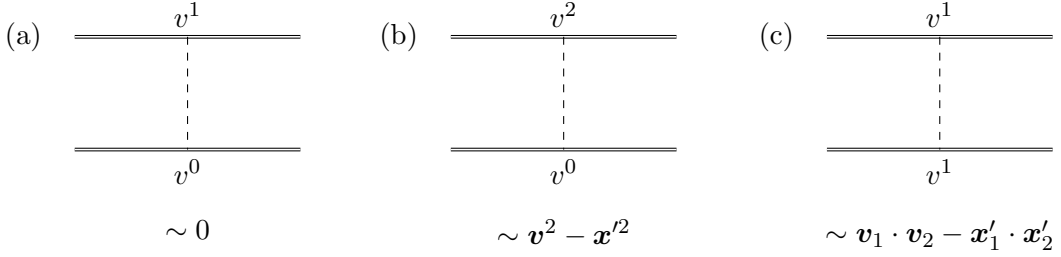
\begin{figure} 
\begin{tikzpicture}\centering
  \begin{feynman}
   \vertex (a1);
   \vertex [left=0.3cm of a1] (a10){(a) };
     \vertex[right=1.5cm of a1] (a2);
    \vertex[right=1.5cm of a2] (a3) ;
     \vertex[below=1.5cm of a1] (b1);
     \vertex[below=1.5cm of a2] (b2);
      \vertex[below=1.5cm of a3] (b3);
       \vertex[below=0.8 cm of b2] (c2){$ \sim 0$};

\vertex[right=2cm of a3] (a4) ;
\vertex[left=0.3 cm of a4] (a40) {(b) };
\vertex[right=1.5cm of a4] (a5) ;
    \vertex[right=1.5cm of a5] (a6) ;
       \vertex[below=1.5cm of a4] (b4);
      \vertex[below=1.5cm of a5] (b5);
       \vertex[below=1.5cm of a6] (b6);
        \vertex[below=0.8 cm of b5] (c5){$  \sim \bm v^2 - \bm x'^2 $};

\vertex[right=2cm of a6] (a7) ;
\vertex[left=0.3 cm of a7] (a70) {(c) };
\vertex[right=1.5cm of a7] (a8) ;
    \vertex[right=1.5cm of a8] (a9) ;
       \vertex[below=1.5cm of a7] (b7);
      \vertex[below=1.5cm of a8] (b8);
       \vertex[below=1.5cm of a9] (b9);
        \vertex[below=0.8 cm of b8] (c8){$  \sim \bm v_1 \cdot \bm v_2 -  \bm x'_1 \cdot \bm x'_2 $};

      \diagram* {
      (a1)-- [double, edge label=$ v^1$ ](a3)  ,
       (b1)-- [double, edge label'=$ v^0$ ](b3)  ,
              (a2)-- [dashed](b2),
         };
\diagram* {
      (a4)-- [double, edge label=$ v^2$](a6)  ,
        (b4)--[double, edge label'=$ v^0$](b6) ,
         (a5)-- [dashed](b5),
                  };        

\diagram* {
      (a7)-- [double, edge label=$ v^1$](a9)  ,
        (b7)--[double, edge label'=$ v^1$](b9) ,
         (a8)-- [dashed](b8),
                  };        
           \end{feynman}
  \end{tikzpicture}
  \caption{Diagram (a) contributes at $\mathcal{O} (v H)$, but it vanishes as shown in Eq.\eqref{eq,V10}. Diagrams (b) and (c) contribute at $\mathcal{O} (v^2 H). $} \label{fig:v2dia}
\end{figure}

To understand the leading order gravitational potential between two strings, it is therefore necessary to expand the string action to higher order. As shown in Fig.\ref{fig:v2dia} (a),  there is only one diagram to consider at  $\mathcal{O}(v H) $ and this diagram vanishes since
\begin{equation}
\label{eq,V10}
  V^{(1)}_{\mu\nu}= -\bm v^a I_{0a, \mu\nu} +\bm  x'^a I_{1a,\mu\nu},\quad   \text{Fig.\ref{fig:v2dia} (a)}\propto V^{(1)}_{\mu\nu} P^{\mu\nu,\alpha\beta} V^{(0)}_{\alpha\beta} = 0 \ .
\end{equation}
At $\mathcal{O}(v^2 H) $, there are two diagrams, shown in Fig.\ref{fig:v2dia} (b) and Fig.\ref{fig:v2dia} (c). To compute the amplitude arising from Fig.\ref{fig:v2dia} (b) we need the Feynman rule for the string-potential graviton vertex at order $v^2$, which is given by $\Gamma^{(2)}_{\mu\nu}$ with
\begin{equation}
  V^{(2)}_{\mu\nu} = -\frac{1}{2}(\bm v^a \bm v^b - \bm x'^a \bm x'^b)I_{ab,\mu\nu}- \frac{1}{2} (\bm v^2 I_{11,\mu\nu} +\bm x'^2 I_{00,\mu\nu})  + (\bm v\cdot \bm x') I_{01,\mu\nu}-
  \frac{2}{8}(\bm v^2 -\bm x'^2)(I_{00,\mu\nu}-I_{11,\mu\nu}) \  .
\end{equation}
The combined contributions of the diagrams give
\begin{align}
  \text{Fig. \ref{fig:v2dia} (b)+(c)}&= \frac{ - T_1 T_2}{ \mpl^2}  \int d^2\sigma_{1} ~d^2\sigma_{2}   \left(V^{(0)}_{\mu\nu}  V^{(2)}_{\alpha\beta} +V^{(1)}_{\mu\nu}  V^{(1)}_{\alpha\beta}\right) \braket{H^{\mu\nu}H^{\alpha\beta}}\nonumber\\
  &=  \frac{ i ~T_1 T_2}{ \mpl^2}  \int d^2\sigma \int_{\bm k} e^{-i \bm k (\bm x_1-  \bm x_2) }\frac{1}{\bm k^2 } \frac{1}{4} \left[(\bm v_1-\bm v_2)^2-(\bm x_1'-\bm x_2')^2 \right]\nonumber\\
  &= -\frac{ i ~T_1 T_2}{8\pi \mpl^2} \int d^2\sigma \log\left(\frac{|\bm x_1- \bm x_2|}{r_\text{IR}}\right) \left[(\bm v_1-\bm v_2)^2-(\bm x_1'-\bm x_2')^2 \right] \ .
\end{align}

Thus, to this order, the effective action reads
\begin{equation}
\label{eq,potentialnonzero}
S^{\text{eff}}_ {L  \frac{\Delta^2}{v^{6}}} = -\int d^2 \sigma\frac{T_1 T_2}{8\pi \mpl^2} \log\left(\frac{|\bm x_1- \bm x_2|}{r_\text{IR}}\right)   \left((\bm v_1-\bm v_2)^2-(\bm x_1'-\bm x_2')^2 \right) \ .
\end{equation}
The negative sign combined with the negative logarithm gives us a negative gravitation potential as in the point particle case. One could use this potential to confirm that the gravitational force between binary strings is attractive, as expected. 

\subsection{EFT for a single radiation graviton}

In this section, we find the contribution to the effective action which contains only one radiation graviton $\bar h$. As indicated below Eq.\eqref{krad}, there are two types of radiation in the binary system. One is purely from the multipole expansion of a single string and has momentum $\bm k \sim {\sqrt{\Delta}}/{r_w}$, while the other one involves gravitational interactions between the strings and has typical momentum scaling as $\bm k \sim {\sqrt{\Delta}}/{r_o}$. Since $r_w \ll r_o$, the leading order of the effective action with a single radiation graviton is the one from a single wiggly string. We start by discussing the radiation from a single string, where $\bar{h}\sim{\sqrt{\Delta}}/{r_w}$.

The leading order term with a single radiation graviton coming from Eq.\eqref{eq:expansion} is,
\begin{equation}
\label{eq,Seffrad1}
S^\text{eff}_{\frac{L}{v^2}\frac{\bar h}{\mpl }} = \int d^2 \sigma\sum_i \frac{T_i}{2 \mpl} \left(\bar{h}_{00}- \bar h_{11} \right) \ .
\end{equation}
Note that, under a gauge transformation $h_{\mu\nu} \to h_{\mu\nu} + \partial_\mu \xi_\nu + \partial_\nu \xi_\mu$, the variation of Eq.\eqref{eq,Seffrad1} 
vanishes after integrating by parts. In fact, our effective action should be gauge invariant order by order, but the proof of this becomes more complex at higher orders.

The next-to-leading order contribution contains terms from two different origins -  those arising from the Lagrangian expanded directly to order $\mathcal{O}(v \bar h)$, and those coming from the multipole expansion around the center mass $\bm x_{CM}$. The effective action obtained from summing these contributions is
\begin{equation}
S^\text{eff}_{\frac{L}{v}\frac{\bar h}{\mpl }} = \int d^2 \sigma \sum_i \frac{T_i}{2 \mpl} \left( 2\bm v_i^a \bar h_{0a} -  2\bm x_i'^a \bar h_{1a} + (\bm x_i^a-\bm x_{CM}^a) \partial_a \bar h_{00} - (\bm x_i^a-\bm x_{CM}^a)  \partial_a \bar h_{11} \right)=0 \ , \label{Lv3/2}
\end{equation}
where in the last equality we  have integrated by parts and used the fact that  $\dot{\bm x}_{CM} = \bm x'_{CM} =0$  in the center of mass frame. Similarly, we find that the next-to-next-to-leading order effective action is
\begin{align}
  \label{eq,lv52}
S^\text{eff}_{L\frac{\bar h}{\mpl }}&= &\int d^2 \sigma  \sum_i \frac{T_i}{2\mpl} \left(  \frac{1}{2}\bm x^a\bm x^b  \partial_a\partial_b(\bar h_{00} - \bar h_{11}) + \bm x^b \left(2\bm v^a \partial_b\bar h_{0a} -  2\bm x'^a \partial_b\bar h_{1a} \right) \right. \nonumber\\
  & & \left. +\bm  v^a \bm v^b \bar h_{ab}- \bm x^{'a} \bm x^{'b} \bar h_{ab} + \frac{1}{2} (\bm v^2 + \bm x^{'2} )( \bar h_{00}+ \bar h_{11}) -2  (\bm v\cdot \bm x') \bar h_{01}  \right)  \ ,
\end{align}
where we have now set $\bm x_{CM}=0$. After integrating by parts, we can express this as
\begin{equation}
S^\text{eff}_{L\frac{\bar h}{\mpl }}=\int d^2 \sigma   \sum_i \frac{T_i}{2\mpl}  \left( -R_{0a0b} + R_{1a1b}\right)  \left(\bm x^a\bm x^b\right)  + \frac{1}{2} (\bm v^2 + \bm x'^2 )( \bar h_{00}+ \bar h_{11}) -2  (\bm v\cdot \bm x') \bar h_{01} \ .  \label{s3}
\end{equation}
Notice that $\text{Tr}(R_{0a0b} -R_{1a1b}) = 0 $, which can be seen by using that $R_{0a0a} -R_{1a1a} =R_{0\mu0\mu} - R_{1\mu1\mu}$, and that $R_{0\mu0\mu}$ and $R_{1\mu1\mu} $ vanish for on-shell gravitons in the harmonic gauge ($\partial_\mu h_{0\mu} = \frac{1}{2} \partial_0 h$). Defining the quadrupole moment,
\begin{equation}
Q_{ab} = \sum_i T_i \left(\bm x^a\bm x^b-\frac{1}{2}\bm  x^2 \delta_{ab}\right)  \ ,
\end{equation}
 we can rewrite Eq.\eqref{s3} as
 \begin{equation}
	 \label{eq,singleradi}
 S^\text{eff}_{L\frac{\bar h}{\mpl }}=\int d^2 \sigma \left[ \frac{Q_{ab}}{2\mpl}  \left( -R_{0a0b} + R_{1a1b}\right)   + \sum_i \frac{T_i}{2\mpl} \left( \frac{1}{2} (\bm v^2 + \bm x'^2 )( \bar h_{00}+ \bar h_{11}) -2  (\bm v\cdot \bm x') \bar h_{01} \right)  \right] \ .
 \end{equation}
 The first term in this expression corresponds to a quadrupole coupling, and will give rise to gravitational waves. Meanwhile, the second term is proportional to the total energy of the string at leading order. Since at this order the energy is a conserved quantity, this coupling will not give rise to radiation. The third term also corresponds to a conserved quantity in time --- in this case it is proportional to the leading order Noether charge corresponding to invariance under translations in $\sigma$. The gauge invariance of this Lagrangian can be checked after integrating by parts and using the conservation of the stress-energy tensor, $\nabla_\mu T^{\mu\nu}=0$, where $T^{\mu\nu} \propto  \int d^2 \sigma (\dot x^\mu \dot x^\nu - x^{\mu'} x^{\nu'}) \sim  \int d^2 \sigma ~x^\mu (\ddot x^\nu - x^{'' \nu}) $.

It is important to note that, from the results we have just obtained, a single wiggly string will itself give rise to gravitational radiation. Contrary to the analogous calculations in the point-particle case, the contributions we found are gauge invariant on their own. Furthermore, there is no radiation arising from gravitational interactions at this order, which implies that we can already self-consistently obtain the radiation arising from a single string using the effective action found above.
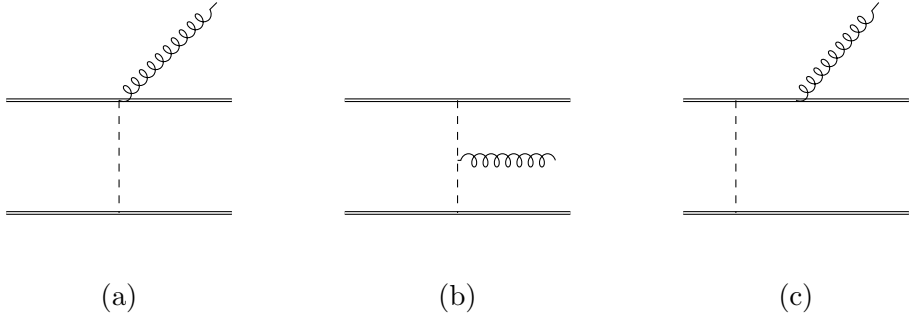
\begin{figure} 
\begin{tikzpicture}\centering
  \begin{feynman}
   \vertex (a1);
     \vertex[right=1.5cm of a1] (a2);
      \vertex[right=1.3cm of a2] (a21);
 \vertex[above=1.3cm of a21] (a22);
    \vertex[right=1.5cm of a2] (a3) ;
     \vertex[below=1.5cm of a2] (b2);
      \vertex[below=1.5cm of a1] (b1);
 \vertex[below=1.5cm of a3] (b3);
   \vertex[below=0.8cm of b2] (c2){(a)};

\vertex[right=1.5cm of a3] (a4) ;
\vertex[right=1.5cm of a4] (a5) ;
\vertex[below=0.8 cm of a5] (a51);
\vertex[right=1.3 cm of a51] (a52);
    \vertex[right=1.5cm of a5] (a6) ;
       \vertex[below=1.5cm of a4] (b4);
      \vertex[below=1.5cm of a5] (b5);
      \vertex[above=0.7cm of b5] (b51);
       \vertex[below=1.5cm of a6] (b6);
        \vertex[below=0.8cm of b5] (c5){(b)};       
       \vertex[right=1.5cm of a6] (a7) ;
\vertex[right=1.5cm of a7] (a8) ;
\vertex[left =0.8 cm of a8] (a81);
\vertex[right=1.1 cm of a8] (a82);
\vertex[above=1.3 cm of a82] (a83);
    \vertex[right=1.5cm of a8] (a9) ;
       \vertex[below=1.5cm of a7] (b7);
      \vertex[below=1.5cm of a8] (b8);
      \vertex[below=1.5cm of a81] (b81);
       \vertex[below=1.5cm of a9] (b9);
        \vertex[below=0.8cm of b8] (c8){(c)};

      \diagram* {
      (a1)-- [double  ](a3)  ,
       (b1)-- [double ](b3)  ,
              (a2)-- [dashed](b2),
               (a2)-- [gluon](a22),
         };
           \diagram* {
      (a4)-- [double ](a6)  ,
        (b4)--[double ](b6) ,
         (a5)-- [dashed](b5),
          (a52)-- [gluon](b51),
                  };        
                  \diagram* {
      (a7)-- [double ](a9)  ,
        (b7)--[double ](b9) ,
         (a81)-- [dashed](b81),
          (a8)-- [gluon](a83),
      };        
        
           \end{feynman}
  \end{tikzpicture}
  \caption{The Feynman diagrams contributing to gravitational radiation from binary strings.} \label{fig:diaselfin}
\end{figure}

We now discuss the effective action with a single radiation graviton arising from gravitational interactions of the binary string as shown in Fig. \ref{fig:diaselfin}. In this setting, we will denote the radiation graviton $\bar{h}_b$, since it arises from binary string interactions and thus $\bar{ h}_b\sim \sqrt{\Delta}/r_o$. 
Recall that $r_w \ll r_o$, so the contribution from gravitational interactions in the binary string is not the same order as the contribution arising from a single string, in fact we expect it to be subleading. To confirm this expectation, we analyze the diagrams in Fig. \ref{fig:diaselfin}.  We first consider the case in which the matter vertex is at $\mathcal{O}(v^0)$: 
\begin{equation}
	\label{eq,dia00}
	\raisebox{-28pt}{
		\begin{tikzpicture}\centering
			\begin{feynman}
				\vertex (a1);
				\vertex[right=1cm of a1] (a2);
				\vertex[right=0.8cm of a2] (a21);
				\vertex[above=0.8cm of a21] (a22);
				\vertex[right=1cm of a2] (a3) ;
				\vertex[below=1cm of a2] (b2);
				\vertex[below=1cm of a1] (b1);
				\vertex[below=1cm of a3] (b3);
				
				\vertex[right=0.5cm of a3] (c3) ;
				\vertex[below=0.25cm of c3] (d3){$=$} ;
				
				\vertex[right=1cm of a3] (a4) ;
				\vertex[right=1cm of a4] (a5) ;
				\vertex[below=0.3 cm of a5] (a51);
				\vertex[right=0.8 cm of a51] (a52);
				\vertex[right=1cm of a5] (a6) ;
				\vertex[below=1cm of a4] (b4);
				\vertex[below=1cm of a5] (b5);
				\vertex[above=0.2cm of b5] (b51);
				\vertex[below=1cm of a6] (b6);
				
				\diagram* {
					(a1)-- [double, edge label = {$v^0$}  ](a3)  ,
					(b1)-- [double ,edge label' = {$v^0$}  ](b3)  ,
					(a2)-- [dashed](b2),
					(a2)-- [gluon](a22),
				};
				\diagram* {
					(a4)-- [double,edge label = {$v^0$}   ](a6)  ,
					(b4)--[double,edge label' = {$v^0$}   ](b6) ,
					(a5)-- [dashed](b5),
					(a52)-- [gluon](b51),
				};        
			\end{feynman}
		\end{tikzpicture}
	}=0
\end{equation}
The effective action arising from these two diagrams is $\mathcal{O}(\frac{L}{v^2}\frac{\bar h_b}{\mpl} \left(\frac{\Delta}{v^4} \right)^2)$. Due to the symmetry of the vertices' tensor structure, the diagrams shown in Eq.\eqref{eq,dia00} automatically vanish. We then compute the next-leading-order matter vertex $\mathcal{O}(v^1)$ and find that each diagram vanishes on its own.
\begin{equation}
	\label{eq,dia01}
	\raisebox{-28pt}{
		\begin{tikzpicture}\centering
			\begin{feynman}
				\vertex (a1);
				\vertex[right=1cm of a1] (a2);
				\vertex[right=0.8cm of a2] (a21);
				\vertex[above=0.8cm of a21] (a22);
				\vertex[right=1cm of a2] (a3) ;
				\vertex[below=1cm of a2] (b2);
				\vertex[below=1cm of a1] (b1);
				\vertex[below=1cm of a3] (b3);
				
				\vertex[right=0.5cm of a3] (c3) ;
				\vertex[below=0.25cm of c3] (d3){$=$} ;
				
				\vertex[right=1cm of a3] (aa1);
				\vertex[right=1cm of aa1] (aa2);
				\vertex[right=0.8cm of aa2] (aa21);
				\vertex[above=0.8cm of aa21] (aa22);
				\vertex[right=1cm of aa2] (aa3) ;
				\vertex[below=1cm of aa2] (bb2);
				\vertex[below=1cm of aa1] (bb1);
				\vertex[below=1cm of aa3] (bb3);
				
				\vertex[right=0.5cm of aa3] (cc3) ;
				\vertex[below=0.25cm of cc3] (dd3){$=$} ;
				
				\vertex[right=1cm of aa3] (a4) ;
				\vertex[right=1cm of a4] (a5) ;
				\vertex[below=0.3 cm of a5] (a51);
				\vertex[right=0.8 cm of a51] (a52);
				\vertex[right=1cm of a5] (a6) ;
				\vertex[below=1cm of a4] (b4);
				\vertex[below=1cm of a5] (b5);
				\vertex[above=0.2cm of b5] (b51);
				\vertex[below=1cm of a6] (b6);
				
				\diagram* {
					(a1)-- [double, edge label = {$v^1$}  ](a3)  ,
					(b1)-- [double ,edge label' = {$v^0$}  ](b3)  ,
					(a2)-- [dashed](b2),
					(a2)-- [gluon](a22),
				};
				\diagram* {
					(aa1)-- [double, edge label = {$v^0$}  ](aa3)  ,
					(bb1)-- [double ,edge label' = {$v^1$}  ](bb3)  ,
					(aa2)-- [dashed](bb2),
					(aa2)-- [gluon](aa22),
				};
				\diagram* {
					(a4)-- [double,edge label = {$v^1$}   ](a6)  ,
					(b4)--[double,edge label' = {$v^0$}   ](b6) ,
					(a5)-- [dashed](b5),
					(a52)-- [gluon](b51),
				};        
			\end{feynman}
		\end{tikzpicture}
	}=0
\end{equation}
Another possible diagram is shown in Fig. \ref{fig:diaselfin} (c), this is the first non-vanishing contribution to the effective action, but it contributes at  $\mathcal{O}(L \frac{\Delta^2}{v^6}  \frac{\bar{h}_b}{\mpl} ) $, which is smaller than the single string contribution computed in the previous section.
 
\subsection{Radiation Power Analysis} \label{secRad}
The preceding analysis shows that Eq.\eqref{eq,singleradi} is the lowest order contribution to the effective action with a single radiation graviton. To complete the derivation of the full EFT, we would now like to integrate out $\bar h$,
\begin{equation}
\exp{\left[i S^{\rm eff}[x^i]\right]} = \int \mathcal{D} \bar{h}_{\mu\nu} \exp{i  S^{\rm eff}[x^i,\bar h]}  \ .
\end{equation}
As in the previous case, we are not going to perform the path integral, and instead will again simply use the matching procedure.

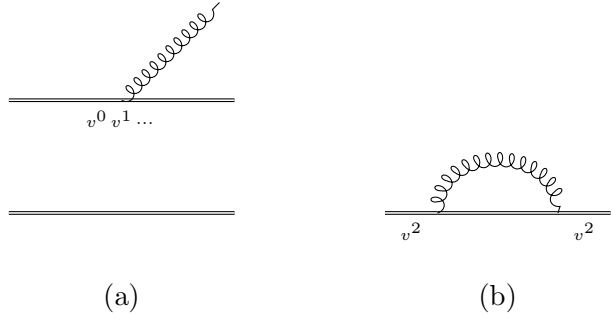
\begin{figure}[!tb]
\begin{tikzpicture}\centering
  \begin{feynman}
   \vertex (a1);
     \vertex[right=1.5cm of a1] (a2);
      \vertex[right=1.3cm of a2] (a21);
 \vertex[above=1.3cm of a21] (a22);
    \vertex[right=1.5cm of a2] (a3) ;
     \vertex[below=1.5cm of a2] (b2);
      \vertex[below=1.5cm of a1] (b1);
 \vertex[below=1.5cm of a3] (b3);
   \vertex[below= 0.8 cm of b2] (c2){(a)};

\vertex[right=2cm of a3] (a4) ;
\vertex[right=1.5cm of a4] (a5) ;
    \vertex[right=1.5cm of a5] (a6) ;
       \vertex[below=1.5cm of a4] (b4);
      \vertex[right =0.7 cm of b4] (b51);
          \vertex[below=1.5cm of a6] (b6);
  \vertex[left=0.7 cm of b6] (b52);
   \vertex[below= 2.3 cm of a5] (c5){(b)};

      \diagram* {
      (a1)-- [double, edge label'=$ \scriptscriptstyle v^0\, v^1\, \cdots $  ](a3)  ,
       (b1)-- [double ](b3)  ,
               (a2)-- [gluon](a22),
         };
\diagram* {
          (b4)--[double, edge label'=$ \scriptscriptstyle v^2  $ ](b51)--[double ](b52)--[double, edge label'=$ \scriptscriptstyle v^2  $ ](b6) ,
           (b51)--[gluon, half left ](b52) ,
           };        
           \end{feynman}
  \end{tikzpicture}
  \caption{Diagram (a) corresponds to the a graviton radiated from a single string, notice that there are no gravitational interactions between strings. Diagram (b) is the self-energy diagram, whose imaginary part is related to the power radiated.} 
	\label{fig:lag1}
\end{figure}

Since we are interested in what gravitational wave detectors may observe, the relevant quantity we would like to compute is the power radiated by the system. The power radiated is related to the imaginary part of the effective action by
\begin{equation}
\label{eq,radiatedpower}
\frac{1}{T} \operatorname{Im}\left[S^{\mathrm{eff}}[x^i]\right]=\frac{1}{2} \int d E d \Omega \frac{d^{2} \Gamma}{d E d \Omega}, \quad d P=E d \Gamma \ ,
\end{equation}
where $T$ is the time over which the detector observes the gravitational radiation. At leading order, we can simply match the scattering amplitude corresponding to the self-energy diagram on the right-hand side of Fig.\ref{fig:lag1}, which involves the string-radiation graviton vertex arising from Eq.\eqref{eq,singleradi}. This leads to
\begin{equation}
   \label{eq,rad1}
S^{\text{eff}}[x]= -\frac{1}{8\mpl^2}    \int d^2 \sigma_1 d^2 \sigma_2  Q_{ab}Q_{cd}  \braket{(R_{0a0b}-R_{1a1b}) (R_{0c0d}-R_{1c1d})} \ .
\end{equation}
Using the graviton correlation function, we find that
\begin{align}
  \label{eq,2ptR0a0b}
  \braket{(R_{0a0b}-R_{1a1b}) (R_{0c0d}-R_{1c1d})} &= \int_k  \frac{-i}{k^2 - i\epsilon} e^{-i k^0 (x_1^0 -x_2^0)+ i k^1 (x_1^1 -x_2^1)}(\eta_{ac}\eta_{bd}+\eta_{ad}\eta_{bc}) \nonumber\\
&\frac{1}{4}\left(\frac{1}{2}k_0^4 - \frac{1}{2} k_0^2 \bm k^2 +\frac{1}{2}k_1^4 + \frac{1}{2} k_1^2 \bm k^2 + \frac{1}{2} k_0^2 k_1^2 + \frac{3}{8} \bm k^4 \right) \ ,
\end{align}
where we have ignored terms proportional to $\eta_{ab}\eta_{cd}$, since $Q_{ab}$ is traceless. In order to obtain the imaginary part of the effective action, we make use of
\begin{equation}
  \text{Im} \frac{1}{k^2 - i \epsilon} =  i \pi \delta(k^2) \ ,
\end{equation}
and the on-shell condition for the radiation graviton in Eq.\eqref{onshell}. These allow us to write the imaginary part of the string effective action as
\begin{equation}
	\text{Im }S^{\text{eff}}[x] = -\frac{3}{512 \mpl^2}    \int \frac{d k^1 d^2 \bm k}{(2\pi)^3} \frac{1}{ \sqrt{k_1^2 + \bm k^2}} (\bm k^2 + 2k_1 ^2)^2  \left|Q_{ab}(k^0,k^1)\right|^2_{k^0 = \sqrt{k_1^2 + \bm k^2}} \ . \label{ImS}
\end{equation}
where $Q(k^0,k^1)$ is the quadrupole in momentum space,
\begin{equation}
  Q(k^0,k^1) = \int d^2 \sigma~ e^{-i k^0 x^0 + i k^1 x^1} Q(x^0,x^1) \ .
\end{equation}
Using Eq.\eqref{eq,radiatedpower}, we finally obtain the power radiated as,
\begin{align}
P&=\frac{2}{T} \frac{1}{4 \pi^{2}}\int dE ~  E^2  \left(-\frac{3}{512 \mpl^2} (E^2 + k_1 ^2)^2  \left|Q_{ab}(E,k^1)\right|^2\right)\nonumber\\
&= -\frac{3 }{1024\pi \mpl^2 } \braket{ \left((\partial_\tau^3 +\partial_\tau \partial_\sigma^2 )Q_{ab}\right)^2} \ ,
\end{align}
where $E = \sqrt{k_1^2 + \bm k^2}$ and angled brackets represent averaging over the observation time $T$. Here, $k_1^{-1}$ denotes the typical wavelength scale of the wiggles. If we were able to detect this gravitational wave signal directly, it would imply a special direction and break cosmological spatial isometry. However, the gravitational waves generated by cosmic strings mainly contribute to the stochastic gravitational wave background thus we do not expect to observe this feature.

\section{The Renormalized string tension and the background metric}
\label{sec,4}
An important feature of the relativistic theory described by the actions in Eqs.\eqref{EHaction} and \eqref{stringaction} is that it gives rise to a classical Renormalization Group (RG) flow. A non-trivial flow is expected due to the singular nature of the string as described by the Nambu-Goto action. This situation is not exclusive to the string action; other examples of classical RG flows can be found in \cite{Goldberger:2004jt,Solodukhin:1997xn,Goldberger:2001tn,deRham:2007mcp}.  In all these  cases, the singular nature of the sources leads to UV divergences\cite{Geroch:1987qn}. This does not imply that our theory is sick; it just signals our ignorance of the structure of the source at small distances. In fact, these UV divergences can be regularized and renormalized by standard quantum field theory methods. In the point-particle case of \cite{Goldberger:2004jt}, the coefficients of higher order operators, which encode the effects of the finite size of the sources, are renormalized at $2$-loops \footnote{Note  that these loops are not real quantum loops as in standard Feynman graphs, since the worldline does not contribute to the Feynman diagram with a propagator. The point-particle is considered to be a classical source. The situation is the same for the string.}. By using dimensional analysis, we can see that codimension $n\geq2$ sources lead to a logarithmic divergence at the $(n-1)$-loop level. Thus, we expect to find this UV divergence at the $1$-loop level for the string living in a $(3+1)$d spacetime. 

In order to renormalize our relativistic string theory we will consider a single string in a linearized gravitational background and compute the loop corrections to its stress-energy tensor. For simplicity, we will restrict  ourselves to the case of a static, straight string. We will also reconstruct the coordinate-space metric, $h_{\mu\nu}(x)$, of the string and verify that at leading order it corresponds to a locally flat spacetime with a global deficit angle \cite{Vilenkin:2000jqa}. 

\begin{figure}[!tb]
\begin{tikzpicture}\centering
  \begin{feynman}
   \vertex (a1);
     \vertex[right=1.5cm of a1] (a2);
    \vertex[right=1.5cm of a2] (a3) ;
     \vertex[below=2 cm of a2] (b2);
      \vertex[below=1.5cm of a1] (b1);
 \vertex[below=1.5cm of a3] (b3);
   \vertex[below= 1 cm of b2] (c2){(a)};

\vertex[right=2cm of a3] (a4) ;
\vertex[right=1.5cm of a4] (a5) ;
 \vertex[left =1.2 cm of a5] (a51);
\vertex[right=1.2 cm of a5] (a52);
    \vertex[right=1.5cm of a5] (a6) ;
       \vertex[below=2 cm of a51] (b51);
   \vertex[below= 3 cm of a5] (c5){(b)};

\vertex[right=2cm of a6] (a7) ;
\vertex[right=1.5cm of a7] (a8) ;
 \vertex[left =1.2 cm of a8] (a81);
\vertex[right=1.2 cm of a8] (a82);
    \vertex[right=1.5cm of a8] (a9) ;
       \vertex[below=1 cm of a8] (b8);
        \vertex[below=2 cm of a8] (b81);
   \vertex[below= 1 cm of b81] (c8){(c)};

      \diagram* {
      (a1)-- [double  ](a3)  ,
               (a2)-- [photon ](b2),
         };
\diagram* {
          (a4)-- [double  ](a6)  ,
           (a51)--[ half right ](a52) ,
            (a51)--[photon ](b51) ,
           }; 
\diagram* {
          (a7)-- [double  ](a9)  ,
           (a81)-- (b8) ,
	 (a82)-- (b8) ,
            (b8)--[photon ](b81) ,
           };                   
           \end{feynman}
  \end{tikzpicture}
  \caption{The divergent piece for $(a): T^{\mu\nu}_{(0)}(k)~, (b)+(c): T^{\mu\nu}_{(1)}(k)$ .The wiggly line denotes the background field $\mathcal{H}$, while the other solid internal lines correspond to the perturbation field $\mathfrak{h}$.} \label{fig:retension}
\end{figure}
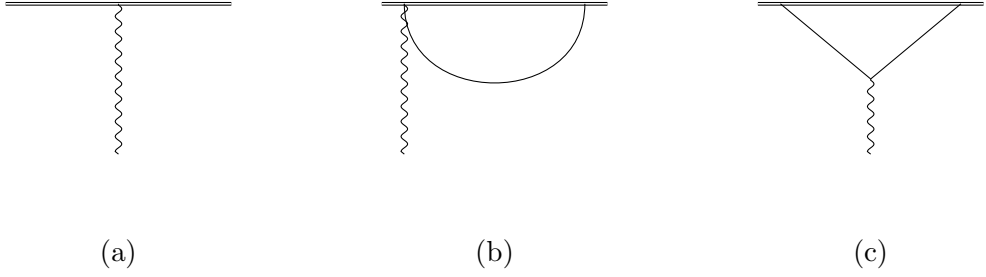

We begin by considering the effective action for the source string in a linearized background, $h_{\mu\nu} = g_{\mu\nu} - \eta_{\mu\nu} $ which can be written as\footnote{The energy-momentum tensor is defined as $T^{\mu\nu} = -2 \partial S/\partial h_{\mu\nu}$. Here we have an additional $- \frac{1}{2}$ factor in the effective action compared with the convention used in \cite{Goldberger:2004jt}.  } 
\begin{equation}
  \label{eq,effS}
  \Gamma[g_{\mu\nu}] = -\frac{1}{2\mpl} \int \frac{d^4 k}{(2\pi)^4} h_{\mu\nu}(-k) (T^{(0)\mu\nu}(k)+ T^{(1)\mu\nu}(k)+T^{(2)\mu\nu}(k)+\cdots)  \  .
\end{equation}
The leading order term corresponds to the original stress-energy tensor of the static, straight string and is given by
 \begin{equation}
   T^{(0)\mu\nu}(k)= T^{\mu\nu}_{\rm string} = T(2\pi)^2 \delta(k^0) \delta(k^1) (u^\mu u^\nu - w^\mu w^\nu)  \ , \label{st}
 \end{equation}
which corresponds to the tree level diagram shown in Fig.\ref{fig:retension}(a). Here, $u^\mu$ and $w^\mu$ are unit vectors in the $x^0$ and $x^1$ directions respectively.

Higher-order terms in Eq.\eqref{eq,effS} correspond to corrections to the stress-energy tensor due to gravitational self-interactions. To include self-interactions of the graviton, it is convenient to use the background field method. We first split the metric perturbation $h_{\mu\nu}$ into two terms,
\begin{equation}
  h_{\mu\nu} = \mathcal{H}_{\mu\nu} +   \mathfrak{h}_{\mu\nu} \ , \label{bck}
\end{equation}
where $\mathcal{H}$ is the (classical) background and $\mathfrak{h}$ is the (quantum) fluctuation. We then proceed to integrate out the quantum degree of freedom to obtain the effective action for the background field 
\begin{align}
  e^{i \Gamma[\mathcal{H}]} &= \int \mathcal{D} [  \mathfrak{h}_{\mu\nu}] ~e^{i S_{EH} [\mathcal{H}+  \mathfrak{h}_{\mu\nu}]  + i S_{GF}[  \mathfrak{h}_{\mu\nu}]  -\frac{i}{2} \int T_{\rm string}^{\mu\nu}(\mathcal{H}_{\mu\nu} +   \mathfrak{h}_{\mu\nu}) }  \nonumber \\
  &=  e^{-\frac{i}{2} \int T_{\rm string}^{\mu\nu}\mathcal{H}_{\mu\nu} }\int \mathcal{D} [  \mathfrak{h}_{\mu\nu}] ~e^{i S_{EH} [\mathcal{H}+  \mathfrak{h}_{\mu\nu}]  + i S_{GF} [  \mathfrak{h}_{\mu\nu}]  -\frac{i}{2} \int T_{\rm string}^{\mu\nu}  \mathfrak{h}_{\mu\nu} }  \ ,
\end{align}
where the gauge fixing term is given by
\begin{equation}
S _{\rm GF} [  \mathfrak{h}] = -\mpl^2 \int d^4 x \sqrt{-\bar g} ~\Gamma_\mu [  \mathfrak{h}] \Gamma^\mu [  \mathfrak{h}]\ ,   \quad  \Gamma_\mu [  \mathfrak{h}] = D_\alpha  \mathfrak{h}^\alpha_{~\mu} - \frac{1}{2} D_\mu  \mathfrak{h}^\alpha_{~\alpha}  \ ,
\end{equation}
with $D_\alpha $ the covariant derivative of the background metric  $\bar g_{\mu\nu} = \eta_{\mu\nu} +\mathcal{H}_{\mu\nu} $. Note that only the quantum field $\mathfrak{h}$ is gauge-fixed, since the background field does not propagate. This ensures that the effective action is gauge invariant and simplifies the computation of the corrections to the stress-energy tensor.  We can now write the effective action as
\begin{equation}
  \Gamma = -  \frac{1}{2}\int T_{\rm string}^{\mu\nu}\mathcal{H}_{\mu\nu}  - i \log{\left(\int \mathcal{D} [ \mathfrak{h}_{\mu\nu}] ~e^{i S_{EH} + i S_{GF} -\frac{i}{2} \int T_{\rm string}^{\mu\nu} \mathfrak{h}_{\mu\nu} }\right)}  \ ,
\end{equation}
which is the one-particle-irreducible (1PI) effective action. The leading order diagrams contributing to the effective action are shown in Fig.\ref{fig:retension}. In the following section, we calculate these diagrams, which give the renormalized energy-momentum tensor with two tension insertions.

\subsection{Renormalized energy-momentum tensor with two tension insertions }
We proceed to compute the correction to the stress-energy tensor from graphs with two tension insertions. At this order, we have two 1PI diagrams (see Fig.\ref{fig:retension}, diagrams (b) and (c)). The two-graviton tension vertex that appears in diagram (b) of Fig.\ref{fig:retension} can be obtained from the term of order $v^0 \mathfrak{h}\mathcal{H}$ in the expansion of the Lagrangian,
\begin{equation}
  \mathcal{V}^{(\mathfrak{h}\mathcal{H})}_{\mu\nu,\alpha\beta} \propto -  I_{\mu\nu,01}I_{\alpha\beta,01} +\frac{1}{4} (I_{\mu\nu,00}+I_{\mu\nu,11})(I_{\alpha\beta,00}+I_{\alpha\beta,11})  \ .
\end{equation}
On the other hand, the Feynman rule for the string-quantum graviton vertex can be obtained in a similar way to Eq.\eqref{eq,vertex1}
\begin{equation}
 \mathcal{\gamma}^{(\mathfrak{h})}_{\alpha\beta} =\frac{i T}{\mpl} (2\pi)^2 \delta(k^0)\delta(k^1)  \mathcal{V}^{(\mathfrak{h})} _{\alpha\beta} \ ,  \quad  \mathcal{V}^{(\mathfrak{h})} _{\alpha\beta} =  \frac{1}{2}  (u^\mu u^\nu-w^\mu w^\nu) I_{\mu\nu,\alpha\beta} \ .
\end{equation}

One can see that the contribution from this diagram vanishes due to the symmetries of the vertices involved, as in Eq.\eqref{eq,V10},
\begin{equation}
 \text{Fig.\ref{fig:retension}~(b)}\propto  \mathcal{V}^{( \mathfrak{h}\mathcal{H})}_{\mu\nu,\alpha\beta} P^{\mu\nu,\rho\sigma}  \mathcal{V}^{(\mathfrak{h})}_{\rho\sigma} = 0 \ .
\end{equation}

Diagram (c) is more involved since it contains the 3-graviton vertex from the Einstein-Hilbert action. After expanding the EH action around a flat metric,
we see that the Lagrangian at order $h^3$ \cite{Berends:1974gk} is
\begin{align}
 \mathcal{L} _{(3)} &= -2h^{\alpha,\mu}_\beta h^{\nu}_\alpha h^{\beta}_{\mu ,\nu} +  h^{\alpha}_\beta h^{\beta,\mu}_\nu h^{\nu}_{\alpha ,\mu}- h^{\alpha}_\beta h^{\mu,\nu}_\alpha h^{\beta}_{\nu ,\mu} - h^{\alpha}_\beta h^{\beta,\mu}_\alpha h^{\nu}_{\nu ,\mu} -\frac{1}{2} h^{\alpha}_{\beta,\mu} h^{\beta,\nu}_\alpha h^{\mu}_{\nu} - h^{\alpha,\mu}_{\alpha,\nu} h^{\beta}_\mu h^{\nu}_{\beta} \nonumber\\
 &+\frac{1}{2} h^{\mu}_{\mu} h^{\alpha}_{\beta,\nu} h^{\nu,\beta}_{\alpha} - \frac{1}{4} h^{\alpha}_{\alpha} h^{\mu,\beta}_\nu h^{\nu}_{\mu,\beta}+\frac{1}{4} h^{\alpha}_{\alpha} h^{\beta,\mu}_\beta h^{\nu}_{\nu,\mu}+\frac{1}{2} h^{\alpha}_{\alpha} h^{\mu,\nu}_{\mu,\beta} h^{\beta}_{\nu} -  h^{\mu}_{\nu} h^{\nu,\alpha}_{\mu,\beta} h^{\beta}_{\alpha} \  ,
\end{align}
where  the graviton should  be split as in Eq.\eqref{bck}. The vertex, $\Gamma_{\mu_1\nu_1,\mu_2\nu_2,\mu\nu} $, arising from the interaction $\mathcal{H}  \mathfrak{h}  \mathfrak{h}$ can be obtained from the expression above.
The gauge fixing term also contributes to the $\mathcal{H}  \mathfrak{h}  \mathfrak{h}$ vertex, but its contribution to the stress tensor ultimately vanishes, as expected, since a gauge fixing term does not affect the physics. 

Having obtained the Feynman rules for all the required vertices, we can now compute diagram (c) in Fig.\ref{fig:retension},
\begin{align}
 & \text{Fig.\ref{fig:retension}~(c)}=\left(\frac{i T}{\mpl}\right)^2  \int_{k_1,k_2}  \Gamma_{\mu_1\nu_1,\mu_2\nu_2,\mu\nu} P^{\mu_1\nu_1,\alpha\beta} P^{\mu_2\nu_2,\rho\sigma} \frac{i^2}{\bm k_1^2 \bm k_2^2} \mathcal{V}^{( \mathfrak{h})}_{\alpha \beta} \mathcal{V}^{( \mathfrak{h})}_{\rho \sigma}   \nonumber\\
  &= -\frac{T^2}{4\mpl^2} (2\pi)^2 \delta(k_3^0) \delta(k_3^1) \left( k_3^2(u_\mu u_\nu - w_\mu w_\nu)\left(1-\frac{5}{2}\epsilon \right)  + 2 (k_{3\mu}k_{3\nu} -  k_3^2 \eta_{\mu\nu})\epsilon\right)I_0 \ ,
\end{align}
where
\begin{equation}
\label{eq,I000}
	I_0 \equiv \int \frac{d^{d-2} \bm{k}_1}{(2\pi)^2} \frac{1}{\bm k_1^2 (\bm k_1 +\bm k_3)^2 }  = -\frac{1}{ 2\pi \bm k_3^2 } \left(\frac{1}{\epsilon}+ \gamma_E +i\pi + \log\left(\frac{\bm k_3^2 }{4\pi}\right)\right)
\end{equation}
is calculated by regularizing the integral via dimensional regularization, as described in Eq.\eqref{dimreg}.

From this result we see that the energy-momentum tensor at 1-loop level is,
\begin{eqnarray}
  T_{\mu\nu}^{(1)} = \frac{T^2}{8\pi \mpl^2}(2\pi)^2 \delta(k_3^0) \delta(k_3^1) &&\left\{(u_{\mu} u_\nu - w_\mu w_\nu) \left(\frac{1}{\epsilon}+  \log\left(\frac{\bm k_3^2}{4\pi\mu^2} \right)+\gamma_E +i \pi\right)\right.\nonumber\\
	&+& \left.2\left(\frac{k_{3\mu} k_{3\nu}}{ k_3^2} - \eta_{\mu\nu} \right) - \frac{5}{2}(u_{\mu} u_\nu - w_\mu w_\nu) \right\} \ ,
\end{eqnarray}
As expected, this is a divergent result that can be made finite by adding appropriate counterterms. In fact, since this correction has the same form as Eq.\eqref{st}, the $1/\epsilon$ divergence can be absorbed by a counter term
\begin{equation}
	T_{CT}^{\mu\nu} =T (2\pi)^2 \delta(k^0)\delta(k^1) \delta_T (u^\mu u^\nu -w^\mu w^\nu ) \ .
\end{equation}
This is equivalent to rewriting the bare string tension as $T=T^r(1+\delta_T)$, where $T^r$ is  the renormalized tension, and neglecting higher order $\delta_T$ terms. Working in the $\overline{\rm{MS}}$ scheme, we then find that the renormalized energy-momentum tensor at 1-loop order is,
\begin{eqnarray}
	\label{renT}
	  T^{\rm{r}}_{ \mu\nu}(\bm k) =&&T^r (2\pi)^2 \delta(k^0)\delta(k^1)  (u^\mu u^\nu -w^\mu w^\nu )\ \nonumber\\
     && -\frac{T^{\rm r2}(\mu)}{8\pi \mpl^2} (2\pi)^2 \delta(k^0)\delta(k^1) \left(2 \left(\eta_{\mu\nu} -\frac{k_\mu k_\nu}{k^2}\right) +(u_{\mu} u_\nu - w_\mu w_\nu)\left( \log \frac{\bm k^2}{\mu^2} -\frac{5}{2} \right) \!\!\right) \ ,
\end{eqnarray}
where the $ \log\left({\bm k^2}/{\mu^2} \right)$ is the expected UV divergence similar to the one found in previous work \cite{Cannella:2008nr,Buonanno:1998kx}. Although we worked in $d = 4+2\epsilon$ dimensions to regulate it, the leftover logarithmic divergence is an ultraviolet one. Notice that the density $\rho = - T^0_{~~0} $ and pressure $p_1 = T^1_{~~1}$  have equal magnitude but opposite sign, so they do not contribute to the Newtonian potential $\nabla^2 \Phi = 4\pi G (\rho + p_1 +p_2 +p_3)$. However, after diagonalizing the spatial directions, one finds that $p_2 + p_3 $ is non-zero. This shows that there is a non-zero Newtonian potential around the static infinitely-long cosmic string at $\mathcal{O}\left(\frac{T^{\rm r2}(\mu)}{8\pi \mpl^2}\right)$ and that the locally flat metric result is only correct at leading order. We will confirm this point in the next section by explicitly calculating the induced metric perturbation and the Ricci scalar.

As usual, physics should not depend on the arbitrary renormalization scale, and thus the effective action is independent of $\mu$. This leads to a classical RG flow for the string's tension given by 
\begin{equation}
\mu\frac{d T^r}{d \mu}=-\frac{(T^r)^2}{4\pi \mpl^2} \ .
\end{equation}

Once we know the RG flow of the tension we can match this EFT coefficient to the expression arising from a microscopic theory of the string, such as that of \cite{Anderson:1997ip},  at the scale $\mu=1/r_s$. We refer the reader to \cite{Rothstein:2003mp} for more details on how this matching procedure should be performed.

\subsection{Spacetime metric generated by  a static, straight string}
From the renormalized stress-energy tensor computed above, we can reconstruct the metric sourced by the string.  We calculate $h_{\mu\nu}(k)$ from our effective action (\ref{eq,effS}) as
\begin{equation}
  h_{\mu\nu}(k) = -\frac{i}{k^2} P_{\mu\nu,\alpha\beta} \frac{i}{\mpl} \frac{\delta\Gamma[g]}{\delta h_{\alpha\beta} (-k)}\Bigg|_{h=0} =  -\frac{1}{k^2} \frac{1}{2\mpl^2}  \left(T^r_{\mu\nu}- \frac{1}{2}\eta_{\mu\nu }{T^r}_\alpha^{~\alpha} \right) \ ,
\end{equation}
where the energy-momentum tensor corresponds to the renormalized one from Eq.\eqref{renT}. Performing an inverse Fourier transform we find at leading order
\begin{align}
  h^{(0)}_{\mu\nu}(x) &=  -\frac{T^r}{2 \mpl^2 }\int \frac{d^4 k}{(2\pi)^4}  \frac{1}{-(k^0)^2 + (k^1) ^2 + \bm k ^2 }(2\pi)^2 \delta(k^0) \delta(k^1)  e^{-i k x} \rm{diag}\{0,0,1,1\}\nonumber\\
  &=  -\frac{T^r}{2 \mpl^2 }\int \frac{d^2 \bm k}{(2\pi)^2}  \frac{1}{ \bm k ^2 } e^{i \bm k\cdot \bm x} \rm{diag}\{0,0,1,1\} \nonumber\\
  &= \frac{T^r}{4\pi \mpl^2} \log{\left(\frac{r}{r_\text{IR}}\right)} \rm{diag}\{0,0,1,1\} \ ,
\end{align}
where $r^2 = |\bm x_a \bm x^a|$ is the radial distance in the plane orthogonal to the string. 
This recovers the locally Minkowski metric around a static straight string upon a coordinate transformation~\cite{Vilenkin:2000jqa}. We can now compute the correction arising from the 1-loop calculation. We find that at this order the metric reads
  \begin{align} \label{hNLO}
  h^{(1)}_{00}(x)& =  - h^{(1)}_{11}(x)=  -\frac{(T^r)^2}{32 \pi^2 \mpl^4} \log \left(\frac{r}{r_\text{IR}} \right)\ , \nonumber \\
  h^{(1)}_{ab}(x) &=  -\frac{(T^r)^2}{32 \pi^2 \mpl^4} \left( \eta_{ab}- \frac{x_a x_b}{|\bm x|^2} + \eta_{ab} \left( \log \left(\frac{rr_{\text{IR}}}{4 e^{-2 \gamma_E} r^2_{\text{UV}}}\right)+5 \right)\log \left(\frac{r }{r_{\text {IR}}}\right)   \right) \ ,
\end{align}
where $r_{\text{UV}} = 1/\mu$ is the UV cutoff; the details of the calculation can be found in Appendix \ref{int}. From this we find that the Ricci scalar at NLO is 
\begin{eqnarray}
  \mathcal{R} _{\text{NLO}}=  \frac{1}{4} h^{(0)} \partial^2  h^{(0)}  - \frac{1}{2} h^{(0)\mu\nu} \partial^2  h^{(0)}_{\mu\nu} - \frac{1}{2}\partial^2  h^{(1)}  = \frac{(T^r)^2}{16 \pi^2 \mpl^4} \frac{1}{r^2 } \ ,
\end{eqnarray}
where $h^{(0)}$ is short for $h^{(0)\mu}_{~~~~\mu}$. This explicitly shows that the metric around the infinitely-long cosmic string is not locally flat at NLO, further confirming the non-vanishing Newtonian potential arising from the spatial part of the stress-energy tensor as discussed in the previous section.

\section{Conclusion}
Effective field theory techniques are playing an increasingly prominent role in our understanding of the gravitational radiation generated by astrophysical sources, such as binary black hole systems. While extensive effort has gone into a detailed understanding of how these methods can be applied to extract large-scale information from point particle sources, less is known about how to extend the techniques to more general settings.

In this paper we have begun such an analysis by focusing on a class of extended objects - cosmic strings - that may well be produced during particle physics phase transitions in the early universe. Cosmic strings can generate gravitational radiation in a variety of ways, with some of the most important and relevant being due to their relativistic motion. Nevertheless, the simplest extension of point-particle techniques is to the more artificial setting in which strings orbit each other in a non-relativistic way and we have shown that even this simple extension of the basic setup requires careful treatment and contains novel features. 

By studying a system of binary infinitely-long cosmic strings, we have seen that the non-relativistic treatment requires an extra expansion parameter, compared to the point particle case. We have derived the relevant power counting rules and the associated generalized virial theorem, and have established the EFT for a binary cosmic string coupled to gravity. We have calculated the leading-order gravitational potential between a pair of wiggly strings and have obtained the gauge-invariant effective action for radiation gravitons order by order. We have then compared the gravitational radiation from a single NR cosmic string with that from the binary cosmic system, for which both the potential graviton and graviton-sourced radiation are included.  

Finally, we have further extended the treatment to the relativistic case of a static straight string, and have analyzed the non-trivial classical RG flow. We obtained the renormalized stress-energy tensor at 1-loop order and showed that the divergences arising in this calculation imply a classical RG flow of the string's tension. Using this result, we have reconstructed the perturbed metric to NLO around a static straight cosmic string, finding that at NLO the metric around the cosmic string is no longer locally flat.

A more realistic setup for the computation of gravitational waves from cosmic strings should involve relativistic cosmic string loops. In this case, we expect that the EFT would also have two expansion parameters, where one could still be the gap between the velocity and bending ($\Delta$), but determining the other one would require a careful analysis of the new setup.

\acknowledgments
We thank Walter Goldberger for useful discussions at an early stage. This work is supported in part by US Department of Energy (HEP) Award DE- SC0013528. M.T. is also supported by NASA ATP grant 80NSSC18K0694, and by the Simons Foundation Origins of the Universe Initiative, grant number 658904. MCG is supported by the STFC grant ST/T000791/1 and the European Union’s Horizon 2020 Research Council grant 724659 MassiveCosmo ERC–2016–COG.

\appendix
\section{Useful integrals} \label{int}
In this appendix, we discuss some important integrals used throughout the paper. In subsection \ref{secRad}, we made use of the following integral
{\small
\begin{equation}
	\int \frac{d^4 k}{(2\pi)^4} \delta((k_0)^2-k_1^2 -\bm k ^2)=   \int \frac{d k^0 d k^1 d^2 \bm k}{(2\pi)^4} \frac{1}{2\sqrt{\bm k^2 + k_1^2}}\left(\delta\left(k^0+\sqrt{\bm k^2 + k_1^2}\right)+\delta\left(k^0-\sqrt{\bm k^2 + k_1^2}\right)\right) \ , \label{onshell}
\end{equation}
}to obtain Eq.\eqref{ImS}.
Afterwards, the integral $I_0$ was introduced in Eq.\eqref{eq,I000}. Here, we use dimensional regularization in $d = 4+2\epsilon$ to compute it
\begin{eqnarray}
I_0&=&  \int \frac{d^{d-2} \bm k }{(2\pi)^{d-2}} \frac{1}{\bm k^2 (\bm k+ \bm p)^2} =  \int \frac{d^{d-2} \bm k }{(2\pi)^{d-2}} \int_0^1 dx  \frac{1}{((1-x)\bm k^2 +x(\bm k+ \bm p)^2)^2}  \nonumber\\
&=&\int_0^1 dx  \int \frac{d^{d-2} \bm k' }{(2\pi)^{d-2}}  \frac{1}{(\bm k'^2 -\Delta)^2} ~,\quad \Delta = (x^2-x)p^2    \nonumber\\
&=& \int_0^1 dx  \frac{1}{(x^2 -x )^{3- d/2}} \frac{1}{(4\pi)^{d/2-1}}\Gamma\left(3- \frac{d}{2}\right) (p^2)^{d/2-3}\nonumber\\
&=& \frac{1}{(4\pi)^{d/2-1}}\Gamma\left(3- \frac{d}{2}\right) (p^2)^{d/2-3} \frac{(-1)^{1+\epsilon}\Gamma(\epsilon)^2}{\Gamma(2\epsilon)}\nonumber\\
&=& -\frac{1}{ 2\pi p^2 } \left(\frac{1}{\epsilon}+ \gamma_E +i\pi + \log\left(\frac{p^2}{4\pi}\right)\right) \ .  \label{dimreg}
\end{eqnarray}
The other relevant integrals correspond to 2d Fourier transformations arising when we compute the metric at NLO in Eq. \eqref{hNLO}. Consider the following integral
\begin{align}
  \int \frac{d^2 \bm k}{(2\pi)^2}  \frac{1}{ \bm k ^2 } e^{i \bm k\cdot \bm x} = \int \frac{k d k d\theta }{(2\pi)^2}  \frac{1}{ k ^2 } e^{i  k  r \cos \theta} = \int \frac{ d k  }{2\pi}  \frac{1}{ k  } J_0(k r) \ ,
\end{align}
which diverges logarithmically at small $k$. Introducing $k_\epsilon$ as an IR cut-off, we have
\begin{equation}
\label{eq,Bessel}
  \int_{k_\epsilon}^\infty d k ~\frac{1}{k} J_0(k r) =  -\gamma_E+ \frac{k_\epsilon^2 r^2 }{8} ~_2F_3\left(1,1; 2,2,2; -\frac{k_\epsilon^2 r^2 }{4}\right)+ \log{\left(\frac{2}{k_\epsilon r }\right)} \ .
\end{equation}
We then expand Eq.\eqref{eq,Bessel} around $k_\epsilon r \ll 1$, and rewrite $k_\epsilon = \frac{1}{\tilde r_{\text{IR}}}$,
\begin{equation}
  \label{eq,I2}
  \int \frac{d^2\bm k}{(2\pi)^2}  \frac{1}{ \bm k ^2 } e^{i \bm k\cdot \bm x}  = -\frac{1}{2\pi} \log{\left(\frac{r}{2 e^{-\gamma_E} \tilde r_\text{IR}}\right)}\equiv -\frac{1}{2\pi} \log{\left(\frac{r}{ r_\text{IR}}\right)} \ ,
\end{equation}
where we absorbed the constants into $ r_\text{IR}$, and defined $r \equiv |\bm x|$. 
Similarly, we have 
\begin{equation}
\label{eq,I1} \frac{1}{|\bm x|^2}\int \frac{d^2 \bm k}{(2\pi)^2} \frac{(\bm k \cdot \bm x)^2}{ (\bm k^2)^2}e^{- i \bm k \cdot \bm x}   = -\frac{1}{4\pi } \left( \log\left(\frac{r}{  r_\text{IR}}\right) + \frac{1}{2}\right) \ .
\end{equation}
Thus we have,
\begin{equation}
\label{eq,I3} \int \frac{d^2 \bm k}{(2\pi)^2} \frac{\bm k_a\bm k_b }{ (\bm k^2)^2}e^{- i \bm k \cdot \bm x}   = \left(-\frac{1}{4\pi } \log\left(\frac{r}{  r_\text{IR}}   \right)+ \frac{1}{8\pi} \right) \eta_{ab} - \frac{1}{4\pi} \frac{\bm x_a \bm x_b}{|\bm x|^2} \ .
\end{equation}
and
\begin{equation}
\label{eq, II}
  \int \frac{d^2\bm k}{(2\pi)^2}  \frac{1}{ \bm k ^2 } \log(\bm k^2) e^{i \bm k\cdot \bm x}  = \int \frac{dk}{2\pi}\frac{1}{k} \log(k^2) J_0(kr) = \frac{1}{2\pi} \left(\log^2\left(\frac{r}{2 e^{-\gamma_E}}\right) - \log^2(\tilde r_\text{IR})\right)
\end{equation}
Notice that in Eq.\eqref{eq, II}, we cannot have a clean expression in terms of ${r}/{  r_\text{IR}}$ as in Eq.\eqref{eq,I2} and \eqref{eq,I1}. With these equations, we are able to obtain the metric perturbation in Eq.\eqref{hNLO}. 

\bibliography{bibliography}

\end{document}